%% file: paper4.tex
\documentclass[english,12pt]{article}
\usepackage{lmodern}
\usepackage[T1]{fontenc}
\usepackage[utf8]{inputenc}
\usepackage[english]{babel}
\usepackage[pdftex]{graphicx}
\usepackage[width=\textwidth]{caption}
\usepackage{amsmath,amsopn,amssymb,bm, dsfont,mathrsfs}
\usepackage[top=1in, bottom=1in, left=1in, right=1in]{geometry}
\usepackage{array,float, placeins,flafter, longtable,booktabs,pdflscape}
\usepackage{verbatim}
\usepackage{natbib}
\usepackage{mathtools}
\usepackage{multirow}
\usepackage{comment}
\usepackage[hidelinks]{hyperref}
\usepackage{booktabs}
\usepackage[flushleft]{threeparttable}
\usepackage{rotating}
\usepackage{xcolor}

\makeatletter

\newtheorem{thm}{Theorem}
\newtheorem{num}{Numerical Result}
\newtheorem{cor}{Corollary}
\newtheorem{prop}{Proposition}
\newtheorem{lem}{Lemma}
\newtheorem{hyp}{Assumption}
\newtheorem{example}{Example}

\newtheorem{counterexample}{Counterexample:}

\newenvironment{exi}[1]{
  \renewcommand\theexample{#1}
  \example
}{\endexample}
\newenvironment{counterexi}[1]{
  \renewcommand\thecounterexample{#1}
  \counterexample
}{\endcounterexample}

\newcommand{\F}{\mathcal{F}}
\newcommand{\R}{\mathbb R}

\newcommand{\ind}[1]{\mathds{1}\left\{#1\right\}}

\newcommand{\Cov}{\text{Cov}}
\newcommand{\eps}{\varepsilon}
\newcommand{\CR}{\text{CR}_{1-\alpha}}

\newcommand{\CI}{\text{CI}_{1-\alpha}}
\newcommand{\CIU}{\text{CI}_{0.95}}
\newcommand{\deriv}[2]{\partial #1/\partial #2}

\newcommand{\indep}{\perp \!\!\! \perp}

\newcommand{\convL}{\stackrel{d}{\longrightarrow}}
\newcommand{\convP}{\stackrel{P}{\longrightarrow}}
\newcommand{\convN}[1]{\stackrel{d}{\longrightarrow} \mathcal{N}\left(#1\right)}
\newcommand{\st}[1]{\texttt{#1}}

\title{Is Inference Conditional on Not Rejecting a Pre-test Less Reliable than Unconditional Inference?\thanks{We would like to thank Henri Fabre for excellent research assistance, the editors Erzo Luttmer and Andres Santos, three anonymous reviewers, Giuseppe Cavaliere, Hannes Leeb, Fedor Petrov, Benedikt P\"otscher, Jonathan Roth, Martin Weidner and participants of various conferences and seminars for their feedback. Chaisemartin was funded by the
European Union (ERC, REALLYCREDIBLE, GA Number 101043899). Views and opinions expressed are
those of the authors and do not reflect those of the European Union or the European Research Council
Executive Agency. Neither the European Union nor the granting authority can be held responsible for
them. \textbf{AI disclosure statement:} ChatGPT and Claude were used to improve the writing of the abstract, introduction, and conclusion. Refine helped us correct some errors in our proofs.}}

\date{}
\begin{document}

\author{Cl\'{e}ment de Chaisemartin\thanks{Sciences Po Paris, clement.dechaisemartin@sciencespo.fr} \and Xavier D'Haultf\oe{}uille\thanks{CREST-ENSAE, xavier.dhaultfoeuille@ensae.fr.}}

\maketitle

\begin{abstract}
Researchers often report an estimate and confidence interval only if a specification test does not reject—for example, only if a pre-trend test supports a difference-in-differences design. Does this make the reported confidence interval less reliable? We show that when the identifying assumptions hold, the answer is no: conditional on passing the pre-test, the confidence interval may over-cover but cannot under-cover. When the assumptions fail, pre-testing can reduce coverage, but under an additional condition it still improves coverage relative to not pre-testing. A calibration to 12 differences-in-differences examples suggests modest effects of pre-testing on coverage.
\end{abstract}

\section{Introduction}

Specification tests are ubiquitous in applied research. To estimate the effect of a treatment, one has to rely on an identifying assumption to recover treated units' outcome without treatment, and researchers typically conduct pre-tests to assess the credibility of that assumption. For example, difference-in-differences (DID) estimation relies on a parallel-trends assumption. Then, researchers usually start by computing a pre-trend estimator, and they may report a DID estimator of, and a confidence interval (CI) for, the average treatment effect on the treated (ATT) only if treated and control units do not experience significantly different outcome trends before treatment. Similarly, in randomized controlled trials (RCTs), if there is attrition or one fears that the randomization may have been compromised, researchers may first conduct balancing tests comparing the average of some pre-determined covariates in the treated and control samples, and they may report the usual difference-in-means estimator and its CI only if covariates are balanced. Analogous pre-tests arise in IV, RDD, and GMM applications. Across all these
settings the structure is the same: an estimator and its CI are computed and
reported conditional on a specification test not having rejected.

\medskip
In principle, pre-testing is a sensible thing to do — it protects against reporting estimates built on a visibly implausible identifying assumption. However, once we condition on the
pretest's outcome, is the CI we go on to report still as reliable as it claims to be? Our main result is surprisingly reassuring: if the identifying assumption is in fact valid, conditioning on passing the pre-test cannot make the usual CI less reliable.

\medskip
To explain our results, let us formalize the problem in a particular case of our more general setup. Suppose a researcher uses an estimator $\widehat{\beta}$ (e.g., a DID estimator) for a scalar parameter $\beta_0$ (e.g., the ATT). Suppose also that under an assumption we call the null hypothesis of valid specification (e.g., parallel trends), $\widehat{\beta}$ is $\sqrt{n}$-consistent and asymptotically normal for $\beta_0$. The usual $(1-\alpha)$-level CI is then
\[
\CI
:=
\left[
\widehat{\beta}-\frac{q_{1-\alpha/2}(\widehat{\beta})}{\sqrt{n}},
\widehat{\beta}+\frac{q_{1-\alpha/2}(\widehat{\beta})}{\sqrt{n}}
\right],
\]
where $q_x(\widehat{\beta})$ is the $x$-th quantile of the asymptotic distribution of $\sqrt{n}(\widehat{\beta}-\beta_0)$. This CI is supposed to \emph{cover} (i.e. contain) $\beta_0$ $(1-\alpha)$\% of the time. And indeed, by construction under the null we have
$$\underset{n\to \infty}{\lim}P(\beta_0\in \CI)=1-\alpha:$$
the \emph{unconditional coverage rate} (UC) of $\CI$ is equal to its \emph{nominal coverage rate} (NC).

\medskip
Now suppose the null has a testable implication: if it holds, some estimator $\widehat{\theta}$ (e.g., a pre-trends estimator) should not be significantly different from zero. The researcher reports $\CI$ only if $\widehat{\theta}$ is not significantly different from zero---that is, only if $\lVert\sqrt{n}\,\widehat{\theta}\rVert\leq q_{1-\alpha}(\widehat{\theta})$. Then, the feature of $\CI$ we actually care about is its \emph{conditional coverage rate} (CC), namely
\[
\mathrm{CC}
:=
\underset{n\to \infty}{\lim} P\!\left(
\beta_0\in \CI
\,\middle|\,
\lVert\sqrt{n}\widehat{\theta}\rVert\leq q_{1-\alpha}(\widehat{\theta})
\right).
\]
However, nothing in the standard toolkit guarantees that CC equals NC: CC could be strictly larger or strictly lower than $1-\alpha$, and the direction of the inequality could depend on the dependence between $\widehat{\theta}$ and $\widehat{\beta}$. One could design a CI that explicitly accounts for pre-testing; our aim instead is to characterize CC of the naive CI above, since that is the one researchers actually report.

\subsubsection*{Results under the null}

Our first main result is that this concern, while valid in principle, cuts in a specific and reassuring direction when the null of valid specification holds. Then, irrespective of the correlation between $\widehat{\beta}$ and $\widehat{\theta}$,
\begin{equation}\label{eq:intro_CC2}
\mathrm{CC}\geq \mathrm{NC}.
\end{equation}
\eqref{eq:intro_CC2} shows that under the null, pre-testing can lead $\CI$ to over-cover, but it can never lead it to under-cover. 
This follows from the fact that, by the Gaussian correlation inequality \citep{royen2014}, if $(Y,X')\sim \mathcal{N}(0,\Sigma)$ and $(c_x,c_y)$ are positive, then
\begin{equation}\label{eq:intro_GCI}
P\!\left(|Y|\leq c_y\,\middle|\,\lVert X\rVert<c_x\right)
\geq
P\!\left(|Y|\leq c_y\right). 
\end{equation}
We also show that this inequality is strict unless $Y$ and $X$ are uncorrelated---a result that appears to be new---which lets us pin down exactly when \eqref{eq:intro_CC2} holds with equality: $\mathrm{CC}=\mathrm{NC}$ if and only if $\widehat{\beta}$ and $\widehat{\theta}$ are asymptotically independent. \eqref{eq:intro_CC2} extends to infinite-dimensional pretests (e.g., Kolmogorov--Smirnov specification tests) and to one-sided inference on $\beta_0$.

\subsubsection*{Results under the alternative}

Then, we investigate what happens when the null of valid specification is violated. Under the alternative, even without any pre-testing $\CI$ does not cover $\beta_0$ $(1-\alpha)$\% of the time, because $\widehat{\beta}$ is biased. Then, the relevant comparison is not just CC against NC, but also CC against UC, the unconditional coverage rate one would get without pre-testing at all: does pre-testing make things worse than not pre-testing?

\medskip
We show two things. First, as explained earlier, if $\widehat{\beta}$ and $\widehat{\theta}$ are not asymptotically independent, 
the inequality in \eqref{eq:intro_CC2} is strict under the null. Then, by continuity, CC continues to exceed both NC and UC in a neighborhood of the null: $\CI$ still over-covers for small violations of the null. Second, under an additional restriction linking the biases of the estimator and the pre-test statistic to their correlation, we show that we always have CC $\geq$ UC, even for very large violations of the null. We give conditions under which this additional restriction holds in the RCT and IV balancing-test settings.

\medskip
On the other hand, we also show that this additional restriction is unlikely to hold in DID settings. When it fails, there exist data generating processes (DGPs) for which CC is below both NC and UC, potentially substantially so. This can happen even for violations of the null small enough that the pre-test often fails to reject. Then, we complement the theory with a calibration exercise: for the twelve published event-study DID papers in the meta-analysis of Roth (2022), we calibrate a differential-linear-trend DGP to each paper's own pre-trend estimate, and compute both UC and CC for the CI on the first post-treatment effect. Averaged across the twelve papers, UC and CC are 81.0\% and 79.2\% respectively. Thus, in this calibration, almost all of the undercoverage comes from the violation of parallel trends itself: conditioning on passing the pre-test lowers coverage by only 1.8 percentage points. Note that this comparison is not entirely fair, because it does not account for the power of the pre-test, which can lead the analyst to correctly reject the null of parallel trends.

\medskip
Taken together, our results paint a more nuanced picture than the common concern that pre-testing mechanically invalidates inference. Under correct specification, conditioning on passing the test cannot induce undercoverage. Under misspecification, it can, but in important cases pre-testing improves coverage relative to reporting the same CI without pre-testing, and in our DID calibration the incremental distortion due to pre-testing is small.

\subsubsection*{Related literature and outline of the paper}

Inference conditional on not rejecting a pre-test has received some attention in the DID literature. Propositions 1 and 4 in \cite{roth2022pretest} show that under parallel trends, the DID estimator remains conditionally unbiased, and that its conditional variance is smaller than its unconditional variance. Neither conditional unbiasedness nor a reduction in conditional variance by itself implies that CC of $\CI$ is weakly larger than its NC; our result directly establishes this property.

\medskip
Our paper is related to, but distinct from, the large literature on inference after model selection \citep[see, e.g.,][and \citeauthor{kuchibhotla2022post}, \citeyear{kuchibhotla2022post}, for a recent survey]{leeb2005model,leeb2006can,berk2013valid, lee2016exact,heller2019post,kuchibhotla2020valid, andrews2024inference}. A simple example of the settings considered by that literature goes as follows. A researcher seeks to estimate $\beta_0$, the coefficient on $X$ in a regression of $Y$ on $X$,
but is unsure whether to also control for $W$. They first regress $Y$ on $(X,W)$, and test whether $\beta_W$, the coefficient on $W$, is zero. If the test does not reject, their CI for $\beta_0$ is based on $\widehat{\beta}^s$, the coefficient from the ``short'' regression of $Y$ on $X$, while if the test rejects their CI is based on $\widehat{\beta}^\ell$, the coefficient from the ``long'' regression of $Y$ on $(X,W)$. The resulting sequential CI is defined regardless of the pretest's outcome---an important difference from our setting, where we take no stance on what happens if the pretest rejects.
Another important difference is that most of this literature considers CIs that account for pre-testing, and selection more generally \cite[see, e.g.][]{lee2016exact,andrews2024inference}, whereas we study the naive CI that ignores pre-testing entirely and ask when it remains valid anyway.
Our paper still relates to some results in this literature: in the linear-regression example above, Proposition 3.1 of \cite{leeb2003finite} shows that under the null that $\beta_W=0$, $\widehat{\beta}^s$ is asymptotically independent of the pre-test, so that CC of $\widehat{\beta}^s$'s naive CI is equal to its NC.   \cite{angelini2024identification} obtain a similar result in the context of proxy-SVARs. This literature has also shown that under local alternatives, UC of sequential CIs can fall well below their NC; we likewise study local alternatives, but our comparison is between CC and UC of the CI attached to $\widehat{\beta}^s$. This is a relevant comparison in our setting, since we do not assume that we have access to an estimator like $\widehat{\beta}^\ell$ that is consistent even under the alternative.

\medskip
The remainder of the paper is organized as follows. Section \ref{sec:null} presents our results under the null of valid specification. Section \ref{sec:alt} presents our results under the alternative, including the calibration exercise. Section \ref{sec:conclu} concludes and discusses what researchers should and should not take away from our paper.

\section{Results under the null hypothesis}
\label{sec:null}

\subsection{Set-up and examples}

We are interested in $\beta_0\in \R^p$. We use an estimator $\widehat{\beta}$,  which is consistent and asymptotically normal if
$(\theta_0,\eta_0)=0$, where $\theta_0\in \R^q$ and $\eta_0\in \R^r$. We assume for now that $\theta_0$ is finite-dimensional, but we consider the infinite-dimensional case in Section \ref{sub:extension_to_testing_infinitely_many_constraints}. Similarly, in all our examples $\eta_0$ is finite-dimensional, but our results would still hold with an infinite-dimensional $\eta_0$. We refer to $(\theta_0,\eta_0)=0$ as the null hypothesis of valid specification. We also have a consistent and asymptotically normal estimator of $\theta_0$, $\widehat{\theta}$. Therefore, $\theta_0=0$ is testable. Accordingly, the null has a testable implication, though it may not be fully testable, unless $\eta_0$ is by definition equal to zero. We review four examples where this set-up applies.

\begin{exi}{DID: set-up.}
\label{ex:DID}
We have a binary treatment $D$, an outcome $Y$, and a panel dataset with $T>2$ periods. Some units (the ``treated'', denoted by $G=1$) receive the treatment at a period $t_0>2$, whereas  other units (``the controls'', denoted by  $G=0$) do not: $D=G\times 1\{t\geq t_0\}$. Letting $Y_t(0)$ and $Y_t(1)$ respectively denote units' potential outcomes at period $t$  without and with treatment, we seek to estimate $$\beta_0:=(E(Y_{t}(1)-Y_{t}(0)|G=1))_{t\in \{t_0,...,T\}},$$ the vector of average treatment effects on the treated (ATTs) at period $t_0$, $t_0+1$, ..., $T$.\footnote{Results also apply if one instead considers $1/(T-t_0+1)\sum_{t=t_0}^T E(Y_{t}(1)-Y_{t}(0)),$ the ATT across all treated units and time periods.} The null hypothesis of valid specification is a parallel-trends assumption: for all $t\geq 2$,
\begin{equation}\label{eq:PT}
E(Y_t(0)-Y_{t-1}(0)|G=1)=E(Y_t(0)-Y_{t-1}(0)|G=0).
\end{equation}
\eqref{eq:PT} is equivalent to
$(\theta_0,\eta_0)=0$,
with
\begin{align*}
\theta_0:=&(E(Y_{t}-Y_{t_0-1}|G=1)-E(Y_{t}-Y_{t_0-1}|G=0))_{t\in \{1,...,t_0-2\}},\\
\eta_0:=&(E(Y_{t}(0)-Y_{t_0-1}(0)|G=1)-E(Y_t(0)-Y_{t_0-1}(0)|G=0))_{t\in \{t_0,...,T\}}.
\end{align*}
$\theta_0=0$ is testable, but $\eta_0=0$ is not.
\end{exi}

\begin{exi}{arguably exogenous: set-up.}
\label{ex:RCT}
We have a binary treatment $D$, an outcome $Y$, and a dataset where some units are treated while others are not. We seek to estimate the ATT, namely $$\beta_0:=E(Y(1)-Y(0)|D=1).$$ $D$ is ``arguably exogenous'': we assume that $D$ and $Y(0)$ are uncorrelated, i.e.
\begin{equation}\label{eq:random}
\eta_{0}:=E(Y(0)|D=1)-E(Y(0)|D=0)=0.
\end{equation}
Under \eqref{eq:random}, the ATT is identified: $\beta_0=E(Y|D=1)-E(Y|D=0)$. However, \eqref{eq:random} does not hold by design, and needs to be argued for.
A usual pre-test assesses if the treatment is uncorrelated with some pre-determined covariates $X_b$:
\begin{equation}\label{eq:balancing}
\theta_{0}:=E(X_b|D=1)-E(X_b|D=0)=0.
\end{equation}
Then, letting $U$ denote the residual from a linear regression of $Y(0)$ on $(1,X_b)$, the null of valid specification could for instance be \eqref{eq:balancing} and
\begin{equation}\label{eq:CIA_strong}
E(U|D=0)-E(U|D=1)=0.
\end{equation}
Together, \eqref{eq:balancing} and \eqref{eq:CIA_strong} imply
$(\theta_0,\eta_0)=0$. Instances where a treatment's exogeneity has to be argued for via balancing tests arise in RCTs, when the randomization protocol may have been compromised or manipulated \citep[see, e.g.,][]{heckman2010analyzing}, or when there is attrition and one wants to argue that treatment is still exogenous in the sample of non-attriters. A similar setup applies to instrumental variable studies, where researchers often use balancing tests to assess their instrument's exogeneity \citep[see, e.g.,][]{angrist1998children}. Finally, a similar setup applies to matching studies, where researchers often use conditional balancing tests to test the conditional independence assumption \citep[see, e.g.,][]{imbens2001estimating}.
\end{exi}

\begin{exi}{RDD: set-up.}
\label{ex:RDD}
We have a binary treatment $D$, an outcome $Y$, and a sharp regression discontinuity design (RDD): $D=1\{R\geq 0\}$, where $R$ is a continuous running variable.\footnote{To simplify the discussion, we focus on sharp RDDs.} We seek to estimate $\beta_0:=E[Y(1)-Y(0)|R=0]$, the average treatment effect (ATE) at the cut-off. If $r\mapsto E[Y(d)|R=r]$ is continuous at 0 for all $d$, $\beta_0=\lim_{r\downarrow 0} E[Y|R=r] - \lim_{r\uparrow 0} E[Y|R=r]$. Two usual pre-tests are a test of continuity of the density of $R$ at 0 \citep{mccrary2008manipulation} and a test that the mean of some pre-determined covariates $X_b$ is continuous at the threshold. The corresponding estimands are
\begin{align*}
\theta_{0,1}:=&\lim_{r\downarrow 0} f_R(r) - \lim_{r\uparrow 0} f_R(r) \\
\theta_{0,2}:=& \lim_{r\downarrow 0} E[X'_b|R=r] - \lim_{r\uparrow 0} E[X'_b|R=r].	
\end{align*}
The null hypothesis of valid specification could for instance be Conditions 1b and 2b in \cite{lee2008randomized}, which imply that $\theta_0:=(\theta_{0,1},\theta_{0,2})$
and
$$\eta_0:=(\lim_{r\downarrow 0} E[Y(0)|R=r] - \lim_{r\uparrow 0} E[Y(0)|R=r],\; \lim_{r\downarrow 0} E[Y(1)|R=r] - \lim_{r\uparrow 0} E[Y(1)|R=r])$$
are both equal to zero.
\end{exi}

\begin{exi}{GMM: set-up.}
\label{ex:gmm}
We seek to estimate $\beta_0=E[h(U,\nu_0)]\in\R^p$ where $U$ is an observed random vector, $h$ is a known function and the parameter $\nu_0\in \R^s$ is unknown. $\nu_0$ is identified and estimated through the moment conditions $E[g(U,\nu_0)]=0$, with $g$ a known function, $g(U,\nu_0)\in \R^q$ and $q > s$. Intuitively, we can only test $q-s$ conditions, since the others are used to identify $\nu_0$. For a given full-rank matrix $A$ of size $(q-s)\times q$, the null hypothesis of valid specification is then $\theta_0:=A E[g(U,\nu_0)]=0$. 
Letting $\eta_0=0$, the null of valid specification is fully testable.
\end{exi}
	
\newcounter{forreset}
\setcounter{forreset}{\arabic{example}}

\medskip
We now introduce the conditions underlying our main result below.

\begin{hyp}~
	\begin{enumerate}
		\item \label{hyp:beta_hat2} $(\theta_0,\eta_0)=0$ and
$$	\left(\widehat{\Sigma}_\beta^{-1/2}\left(\widehat{\beta}-\beta_0\right), \widehat{\Sigma}_\theta^{-1/2}\widehat{\theta} \right) \convN{0,\Sigma}, \;\text{where } \Sigma:=\begin{pmatrix} I_p & \Sigma_{12} \\ \Sigma_{12}' & \Sigma_{22} \end{pmatrix},$$
$I_p$ is the identity matrix of size $p$, $\Sigma_{12}$ is a matrix of size $p\times q$ and $\widehat{\Sigma}_\beta$ (resp. $\widehat{\Sigma}_\theta$ and $\Sigma_{22}$) is positive definite of size $p\times p$ (resp. $q\times q$).
		\item \label{hyp:test2} We consider $J$ tests of $\theta_0=0$, based on the  statistics $(T_{1,n},...,T_{J,n}) \in\R^J$ that satisfy, if $\theta_0=0$,
		\begin{equation}
		T_{j,n} = T_j\left(\widehat{\Sigma}_\theta^{-1/2}\widehat{\theta}\right)+o_P(1).	
			\label{eq:Tn2}
		\end{equation}
		Moreover, $T_j$ is convex on $\R^q$ and satisfies $T_j(-x)=T_j(x)\ge 0$ for all $x\in \R^q$ and $T_j(0)=0$. The critical region of the $j$-th test is $\{T_{j,n}> q_{j,n}\}$, where $q_{j,n}$ is a random variable satisfying $q_{j,n}\convP q_j>0$.
	\end{enumerate}
	\label{hyp:setup}
\end{hyp}
Assumption \ref{hyp:setup}.\ref{hyp:beta_hat2} requires that $(\theta_0,\eta_0)=0$ and $\widehat{\beta}$ is consistent and asymptotically normal for $\beta_0$. Note that $\widehat{\beta}$ could still be consistent for $\beta_0$ even if $(\theta_0,\eta_0)\ne 0$. For instance, in the DID, arguably exogenous, and RDD examples, it is sufficient to have $\eta_0=0$. Importantly, Assumption \ref{hyp:setup}.\ref{hyp:beta_hat2} does not restrict the asymptotic dependence between $\widehat{\beta}$ and $\widehat{\theta}$. It also allows for estimators converging at rates lower than $n^{1/2}$, as is the case in the RDD example. The $J$ different tests in Assumption \ref{hyp:setup}.\ref{hyp:test2} reflect the fact that one often considers different pre-tests. The functions $T_1,...,T_J$ in these pre-tests have to be convex and symmetric around zero: this condition is key to obtaining our main result. On the other hand, we do not require the pre-tests to be consistent or to reach their nominal level asymptotically. The condition $q_j>0$ in Assumption \ref{hyp:setup}.\ref{hyp:test2} ensures that under the null, the probability to reject does not tend to one.

\medskip
In our four leading examples, Assumption \ref{hyp:setup} holds under regularity conditions.

\begin{exi}{DID: conditions under which Assumption \ref{hyp:setup} holds.}
Suppose that we have a sample of $n$ i.i.d. vectors $U_i:=(G_i,Y_{i,1},...,Y_{i,T})'$ and $U_i$ admits a second moment. For $d\in\{0,1\}$, let $n_d:=|\{i:G_i=d\}|$ (with $|A|$ denoting the cardinality of set $A$) and
\begin{align*}
\widehat{\beta}:=&\left(\frac{1}{n_1}\sum_{i:G_i=1}(Y_{i,t}-Y_{i,t_0-1})-\frac{1}{n_0}\sum_{i:G_i=0}(Y_{i,t}-Y_{i,t_0-1})\right)_{t\in \{t_0,...,T\}}\\
\widehat{\theta}:=&\left(\frac{1}{n_1}\sum_{i:G_i=1}(Y_{i,t}-Y_{i,t_0-1})-\frac{1}{n_0}\sum_{i:G_i=0}(Y_{i,t}-Y_{i,t_0-1})\right)_{t\in \{1,...,q\}},
\end{align*}
where $q:=t_0-2$. Let also $\widehat{V}_d$ be the empirical variance of $(Y_{i,t_0}-Y_{i,t_0-1},...,Y_{i,T}-Y_{i,t_0-1})_{i:G_i=d}$ and define $\widehat{\Sigma}_{\beta}:=\widehat{V}_1/n_1 + \widehat{V}_0/n_0$; define $\widehat{\Sigma}^1_{\theta}$ similarly. Under \eqref{eq:PT}, it readily follows from the central limit theorem and Slutsky's theorem that Assumption \ref{hyp:setup}.\ref{hyp:beta_hat2}  holds. To test the null, one can use a so-called $F$-test for pre-trends. This corresponds to $J=1$, $\widehat{\Sigma}_{\theta}=\widehat{\Sigma}_{\theta}^1$, $T_1(x)=x'x$, $T_{1,n}=T_1(\widehat{\Sigma}_\theta^{-1/2}\widehat{\theta})$, and $q_{1,n}$ equal to a quantile of a chi-squared distribution with $q$ degrees of freedom. Assumption \ref{hyp:setup}.\ref{hyp:test2} holds with these definitions. Alternatively, the so-called ``sup-$t$'' test proposed by \cite{montiel2019simultaneous} corresponds to $J=1$, $\widehat{\Sigma}_{\theta}$ equal to the diagonal matrix with diagonal equal to that of $\widehat{\Sigma}_{\theta}^1$, $T_1(x_1,...,x_{q})=\max(|x_1|,...,|x_q|)$, $T_{1,n}=T_1(\widehat{\Sigma}_\theta^{-1/2}\widehat{\theta})$, and $q_{1,n}$ equal to a quantile of $\max(|X_1|/\widehat{\Sigma}^{1/2}_{\theta,1,1},..., |X_q|/\widehat{\Sigma}^{1/2}_{\theta,q,q})$, where $(X_1,...,X_q)\sim \mathcal{N}(0,\widehat{\Sigma}^1_{\theta})$. Again, Assumption \ref{hyp:setup}.\ref{hyp:test2} holds with these definitions.
\end{exi}

\begin{exi}{arguably exogenous: conditions under which Assumption \ref{hyp:setup} holds.}
Suppose that we have a sample of $n$ i.i.d. vectors $U_i:=(D_i,Y_i,X_{b,i}')'$ and  $U_i$ admits a second moment. For $d\in\{0,1\}$, let $n_d:=|\{i:D_i=d\}|$ and
\begin{align*}
\widehat{\beta}:=& \frac{1}{n_1}\sum_{i:D_i=1}Y_i-\frac{1}{n_0}\sum_{i:D_i=0}Y_i\\
\widehat{\theta}:=&\frac{1}{n_1}\sum_{i:D_i=1}X_{b,i}-\frac{1}{n_0}\sum_{i:D_i=0}X_{b,i}.
\end{align*}
Let also $\widehat{V}_d$ be the empirical variance of $(Y_i)_{i:D_i=d}$ and define $\widehat{\Sigma}_{\beta}:=\widehat{V}_1/n_1 + \widehat{V}_0/n_0$; define $\widehat{\Sigma}_{\theta}$ similarly. Then, under \eqref{eq:random}-\eqref{eq:balancing}, the central limit theorem and Slutsky's theorem imply that Assumption \ref{hyp:setup}.\ref{hyp:beta_hat2} holds. To test $\theta_0=0$, one can for instance use an $F$-test. This corresponds to $J=1$, $T_1(x)=x'x$, $T_{1,n}=T_1(\widehat{\Sigma}_\theta^{-1/2} \widehat{\theta})$, and $q_{1,n}$ equal to a quantile of a chi-squared distribution with $q$ degrees of freedom. Assumption \ref{hyp:setup}.\ref{hyp:test2} holds with these definitions.
\end{exi}

\begin{exi}{RDD: conditions under which Assumption \ref{hyp:setup} holds.}
Suppose that we have a sample of $n$ i.i.d. vectors $U_i:=(R_i,Y_{i},X_{b,i}')'$. Let  $\widehat{\beta}$ and $\widehat{\theta}_2$ be respectively the bias-corrected estimators of $\lim_{r\downarrow 0} E[Y|R=r] - \lim_{r\uparrow 0} E[Y|R=r]$ and $\lim_{r\downarrow 0} E[X'_b|R=r] - \lim_{r\uparrow 0} E[X'_b|R=r]$ proposed by \cite*{calonico2014robust}, and $\widehat{\theta}_1$ be the estimator of $\lim_{r\downarrow 0} f_R(r) - \lim_{r\uparrow 0} f_R(r)$ proposed by \cite*{cattaneo2020simple}. Then, under the null of valid specification and the technical conditions in these two papers, one can show that Assumption \ref{hyp:setup}.\ref{hyp:beta_hat2} holds. Researchers often conduct the two pre-tests separately, without accounting for their covariances. Then, $J=2$, $\widehat{\Sigma}_{\theta}$ is diagonal with $j$-th diagonal term equal to  $1/n^{4/5}$ times a consistent estimator of the asymptotic variance of $\widehat{\theta}_j$, $T_j(x)=x_j'x_j$, $T_{j,n}=T_j(\widehat{\Sigma}_\theta^{-1/2}\widehat{\theta})$ and $q_{j,n}$ is a quantile of a chi-squared distribution with one degree of freedom. Assumption \ref{hyp:setup}.\ref{hyp:test2} holds with these definitions.
\end{exi}

\begin{exi}{GMM: conditions under which Assumption \ref{hyp:setup} holds.}
We observe a sample of $n$ i.i.d. vectors $U_i$. Assume that $V(g(U,\nu_0))$ is nonsingular, $J_h:=E[\deriv{h}{\nu}(U,\nu_0)]$ and $J_g:=E[\deriv{g}{\nu}(U,\nu_0)]$ exist, and $J_g$ has full rank. A first possibility to estimate $\nu_0$ is to use a GMM estimator $\widehat{\nu}_W$, obtained using a positive semidefinite weighting matrix $W$.\footnote{There is (almost) no loss of generality to consider a fixed weighting matrix $W$: the GMM estimator with a stochastic matrix $W_n$ converging in probability towards $W$ is asymptotically equivalent to $\widehat{\nu}_W$.} Under regularity conditions, one has \citep{hansen1982large}:
\begin{equation}\label{eq:GMM_asnormnu}
\sqrt{n}(\widehat{\nu}_W - \nu_0) = \frac{-1}{\sqrt{n}}\sum_{i=1}^n (J_g' W J_g)^{-1}J_g' W g(U_i,\nu_0) +o_P(1).
\end{equation}
Then, letting $\widehat{\beta}:=\sum_{i=1}^n h(U_i,\widehat{\nu}_W)/n$, we can show, still under regularity conditions, that
\begin{equation}\label{eq:GMM_asnormbeta}
\sqrt{n}(\widehat{\beta} - \beta_0) = \frac{1}{\sqrt{n}}\sum_{i=1}^n\left[h(U_i,\nu_0)-\beta_0-J_h(J_g'WJ_g)^{-1}J_g'W g(U_i,\nu_0)\right] +o_P(1).
\end{equation}
Recall that $\theta_0= AE[g(U,\nu_0)]$ for a certain matrix $A$. We now show that the $J$-test \citep{sargan1958estimation,hansen1982large} corresponds to choosing a particular $A$. To define it, let $W^*:=V(g(U,\nu_0))^{-1}$ denote the optimal weighting matrix,\footnote{Again, considering the unfeasible optimal GMM estimator is without loss of generality, because the feasible two-step GMM estimator is asymptotically equivalent to $\widehat{\nu}_{W^*}$.} $\Pi:=I_q - W^{*1/2}J_g(J_g'W^*J_g)^{-1} J_g'W^{*1/2}$ and $P$ be an orthogonal matrix such that $\Pi=P\Delta P'$, where $\Delta$ is a diagonal matrix with first $q-s$ ones and then $s$ zeros. Finally, let $A:=[I_{q-s},0_s]\times P'$. Then, let $\widehat{\theta}:= AW^{*1/2}\sum_{i=1}^ng(U_i,\widehat{\nu}_{W^*})/n$. Under regularity conditions, one has
\begin{equation}\label{eq:GMM_asnormtheta}
\sqrt{n}\,\widehat{\theta} = \frac{1}{\sqrt{n}}\sum_{i=1}^n A W^{*1/2}g(U_i,\nu_0) + o_P(1).
\end{equation}
Equations \eqref{eq:GMM_asnormbeta}-\eqref{eq:GMM_asnormtheta} and the central limit theorem imply that Assumption \ref{hyp:setup}.\ref{hyp:beta_hat2} holds, with $\widehat{\Sigma}_\theta=\Sigma_{22}=I_{q-s}$. The $J$-statistic corresponds to $T_1(x)=x'x$, $T_{1,n}=T_1(\sqrt{n}\widehat{\theta})$ and $q_{1,n}$ equal to a quantile of a chi-squared distribution with $q-s$ degrees of freedom. A second possibility to estimate $\beta_0$ and do the test goes as follows. One uses $s$ of the $q$ moment conditions in $E[g(U,\nu_0)]=0$ to estimate $\nu_0$ and $\beta_0$, and one uses the remaining moments to test the null. If the remaining moments are the last ones, say, this corresponds to choosing a $W$ that puts zero weight on the last $q-s$ moments and $A=[0_s,I_{q-s}]$. Then, $\widehat{\theta}:= A\sum_{i=1}^ng(U_i,\widehat{\nu}_W)/n$. Again, one can show that Assumption \ref{hyp:setup} holds under regularity conditions.
\end{exi}

\begin{counterexi}{moment inequalities.}
	Instead of the standard GMM framework above, consider moment inequalities
\begin{equation}
\theta_0:=E[g(U,\nu_0)]\le 0,	
	\label{eq:ineq_mom}
\end{equation}	
see, e.g., \cite{chernozhukov2007estimation}. Then, the null hypothesis would be $\theta_0\le 0$, which does not correspond to our definition. We could redefine $\theta_0$ as $\theta_0=\max(E[g(U,\nu_0)], 0)$. With that definition, the null is $\theta_0=0$, but Assumption \ref{hyp:setup}.\ref{hyp:beta_hat2} does not hold anymore, because of the maximum operator. An important special case of \eqref{eq:ineq_mom} are models with a binary instrument and treatment. Then, \cite{balke1997bounds} show that the assumptions of \cite{imbens1994identification} have a testable implication, which can be cast as \eqref{eq:ineq_mom}. This pre-test does not fall in our setting. While in general, our results do not apply to moment inequality models, we can still handle the case of a single inequality ($q=1$), as explained in Section \ref{ssub:one_sided} below.
\end{counterexi}

\subsection{Main results} 
\label{sec:results}

\subsubsection{Conservative inference}

Our first result provides a lower bound on the asymptotic probability that $\widehat{\Sigma}_\beta^{-1/2}\left(\widehat{\beta}-\beta_0\right)$ belongs to a convex set, conditional on not rejecting the pre-tests.
\begin{thm}
    If Assumption \ref{hyp:setup} holds, for any convex set $C$ that is centro-symmetric (namely, symmetric around the origin),
	\begin{equation}
\lim_{n\to\infty} P\left[\widehat{\Sigma}_\beta^{-1/2}\left(\widehat{\beta}-\beta_0\right) \in C \big| T_{1,n} \le q_{1,n},...,T_{J,n}\le q_{J,n}\right] \ge  P(Z_1 \in C),
		\label{eq:conv_prob_condit_A1}
	\end{equation}
where $Z_1\sim\mathcal{N}(0,I_p)$.
\label{thm:valid_post}
\end{thm}
Heuristically, the proof goes as follows. First, by convexity and symmetry of the $(T_j)_{j=1,...,J}$, the event $\{T_{1,n} \le q_{1,n},...,T_{J,n}\le q_{J,n}\}$ is asymptotically equivalent to $\widehat{\Sigma}_{\theta}^{-1/2}\widehat{\theta}\in C'$, for some centro-symmetric convex set $C'$. Second, under the null of valid specification, $\left(\widehat{\Sigma}_\beta^{-1/2}\left(\widehat{\beta}-\beta_0\right)\right.$, $\left.\widehat{\Sigma}_\theta^{-1/2}\widehat{\theta}\right)\convL (Z_1,Z_2)$, where $Z_2$ is such that $(Z_1, Z_2)\sim \mathcal{N}(0,\Sigma)$. Then,
\begin{align*}
P\left(\widehat{\Sigma}_\beta^{-1/2}\left(\widehat{\beta}-\beta_0\right)\in C, \widehat{\Sigma}_\theta^{-1/2}\widehat{\theta}\in C'\right)
\approx & P(Z_1 \in C, Z_2 \in C') \ge  P(Z_1\in C) P(Z_2 \in C'),
\end{align*}
where the inequality follows from the Gaussian correlation inequality \citep{royen2014}. Finally, the result follows from dividing both sides of the previous display by $P\left(\widehat{\Sigma}_\theta^{-1/2}\widehat{\theta}\in C'\right)\approx P(Z_2 \in C')$, the probability that the pre-test is not rejected.

\medskip
We now turn to the implications of Theorem \ref{thm:valid_post} for (conditional) inference on $\beta_0$. We consider the usual $F$-test of $\beta_0=b_0$ with test statistic $F_n(b_0):=(\widehat{\beta}-b_0)'\widehat{\Sigma}_{\beta}^{-1}(\widehat{\beta}-b_0)$ and critical region $\{F_n(b_0)>q_{1-\alpha}(p)\}$, with $q_{1-\alpha}(p)$ the quantile of order $1-\alpha$ of a chi-squared distribution with $p$ degrees of freedom. We also consider the standard confidence region (CR)
\begin{equation}
\CR=\left\{b\in\R^p: (\widehat{\beta}-b)'\widehat{\Sigma}_{\beta}^{-1}(\widehat{\beta}-b) \le q_{1-\alpha}(p)\right\}.	
	\label{eq:def_CR}
\end{equation}

\begin{cor}\label{cor:CR_tests}
If Assumption \ref{hyp:setup} holds, then:
\begin{enumerate}
	\item If $\beta_0=b_0$, $\lim_{n\to\infty} P\left(F_n(b_0)>q_{1-\alpha}(p)| T_{1,n} \le q_{1,n},...,T_{J,n}\le q_{J,n}\right)\le \alpha$.
	\item $\lim_{n\to\infty} P\left(\beta_0\in\CR| T_{1,n} \le q_{1,n},...,T_{J,n}\le q_{J,n}\right)\ge 1-\alpha$.
\end{enumerate}
\end{cor}
Hence, under Assumption \ref{hyp:setup}, conditional on not rejecting the pre-tests, the usual $F$-tests and CRs are asymptotically conservative. When $\beta_0$ is a scalar ($p=1$), this implies that two-sided tests and CIs on $\beta_0$ are asymptotically conservative. We have focused for simplicity on $F$-tests and standard CRs but given that Theorem \ref{thm:valid_post} applies to any centro-symmetric convex set, the same result as Corollary \ref{cor:CR_tests} also holds for sup-$t$ tests on $\beta_0$ and cartesian products of symmetric two-sided confidence intervals.



\subsubsection{Necessary and sufficient condition for exact inference.} 
\label{ssub:conditions_for_exact_inference}

One may wonder when conditional inference is actually not conservative. An obvious sufficient condition is asymptotic independence between the estimator and the pre-test ($\Sigma_{12}=0$), but is this condition also necessary? The answer is affirmative when we focus on two important types of pre-tests, namely $F$-tests and sup-$t$ tests.

\begin{prop}
	Suppose that Assumption \ref{hyp:setup} holds with $J=1$ and either (i) $T_1(x)=\max(|x_1|,$ $...,|x_q|)$ and $p=1$ or (ii) $T_1(x)=x'x$ and $\Sigma_{22}=I_q$. If \eqref{eq:conv_prob_condit_A1} holds with equality for some compact set $C$ such that $P(Z_1 \in C)>0$, or if Point 1 or 2 of Corollary \ref{cor:CR_tests} hold with equality for some $\alpha\in(0,1)$, then $\Sigma_{12}=0$.
	\label{prop:conserv}
\end{prop}

The result under (i) readily follows from \cite{royen2014}. In case (ii), the result is new, to the best of our knowledge. To prove it, we show in particular that asymptotically,
$$f\mapsto P\left[\widehat{\Sigma}_\beta^{-1/2}\left(\widehat{\beta}-\beta_0\right) \in C \big|T_{1,n}=f\right]$$
is non-increasing and non-constant when $\Sigma_{12}\ne 0$. This follows from a convenient representation of $(Z_1,T_1(Z_2))$, with $(Z_1,Z_2)\sim\mathcal{N}(0,\Sigma)$ (see Equation \eqref{eq:repres} in the proof) and Anderson's theorem \citep{anderson1955integral}. Note that we impose $\Sigma_{22}=I_q$ in (ii). This is a weak condition: it holds whenever a consistent estimator of the asymptotic variance of $\widehat{\theta}$ is available.

\begin{exi}{DID: application of Proposition \ref{prop:conserv}.}
Assume for simplicity that treatment effects are constant ($(Y_t(1)-Y_t(0))_{t\in\{t_0,...,T\}}=\beta_0$) and $Y_{i,t}(0)$ follows a two-way fixed effects models where the error follows a stationary AR(1):
\begin{align}\label{eq:TWFE_AR(1)}
&Y_{i,t}(0)=\alpha_i+\gamma_t +\eps_{i,t},~\eps_{i,t}=\rho \eps_{i,t-1}+u_{i,t},
\end{align}
where the $(\gamma_t)_{t\ge 1}$ are non-stochastic, $u_{i,t}$ is independent of $G_i$, mean zero, and i.i.d. across $i$ and $t$, and where $|\rho| < 1$. Then, some algebra shows that the asymptotic covariance $C_{\ell,t}$ between $\widehat{\beta}_\ell$ and $\widehat{\theta}_t$ is
\begin{align*}
C_{\ell,t} = & \sum_{g=0}^1 \frac{1}{P(G=g)}\Cov(Y_{t_0+\ell-1}(0)-Y_{t_0-1}(0),Y_t(0)-Y_{t_0-1}(0)|G=g) \\
= &  \left(\frac{1}{P(G=1)}+\frac{1}{P(G=0)}\right)V(\eps_{t_0-1})(1-\rho^{\ell})(1-\rho^{t_0-1-t})>0.
\end{align*}
On the other hand, if $\rho=1$, we obtain $C_{\ell,t}=0$. Therefore, unless $\eps_{i,t}$ is a random walk, $\Sigma_{12}\ne 0$, and inference can never be exact.
\end{exi}

\begin{exi}{GMM: application of Proposition \ref{prop:conserv}.}
If $h(U_i,\nu_0)$ does not depend on $U_i$ (e.g., $\beta_0$ is a subvector of $\nu_0$) and $W=W^*$, it follows from \eqref{eq:GMM_asnormbeta} and \eqref{eq:GMM_asnormtheta} that $\Sigma_{12}=0$: using the optimal GMM estimator guarantees exact inference under the null. On the other hand, we have $\Sigma_{12}\ne 0$ in general if either (i) $h(U_i,\nu_0)$ does depend on $U_i$, as is for instance the case in demand estimation, see, e.g., \cite{berry1995automobile}; or (ii) one does not use the optimal GMM estimator, for instance for fear that it exhibits poor finite-sample properties; or (iii) if instead of using GMM, one uses $s$ of the $q$ moment conditions in $E[g(U,\nu_0)]=0$ to estimate $\nu_0$ and the remaining $q-s$ moments to test the null. In all these cases, conditional inference on $\beta_0$ is conservative in general.
\end{exi}


\subsubsection{One-sided tests and CIs.} 
\label{ssub:one_sided}

Assume that $p=1$. Then, one-sided tests and CIs are not covered by Corollary \ref{cor:CR_tests}. Yet it follows from Lemma \ref{lem:GCI} below, a consequence of the Gaussian correlation inequality,\footnote{We thank Fedor Petrov for giving us the proof of this result.} that a similar result holds for such tests and CIs.
\begin{lem}\label{lem:GCI}
	Let $\mu$ denote a mean zero Gaussian measure on $\R^{q+1}$, $K=(-\infty, a]\times \R^q$ with $a\ge 0$ and $L=\R\times C$ where $C\subset \R^q$ is a centro-symmetric convex set. Then $\mu(K\cap L)\ge \mu(K)\mu(L)$. The same result holds with $K=[-a,\infty)\times \R^q$.
\end{lem}
We consider the test statistic $T_n(b_0):=(\widehat{\beta}-b_0)/\widehat{\Sigma}_{\beta}^{1/2}$. To test the null that $\beta_0\leq b_0$, we consider the critical region $\{T_n(b_0)>z_{1-\alpha}\}$, with $z_{1-\alpha}$ the quantile of order $1-\alpha$ of a standard normal distribution. Accordingly, we consider the unilateral CI $(-\infty, \widehat{\beta} + \widehat{\Sigma}_{\beta}^{1/2} z_{1-\alpha}]$. The result below still holds if one considers instead the critical region $\{T_n(b_0)<z_{\alpha}\}$, to test the null that $\beta_0\geq b_0$, or if we consider the unilateral CI $[\widehat{\beta}+\widehat{\Sigma}_{\beta}^{1/2} z_{\alpha},+\infty)$.
\begin{prop}\label{prop:unilat_inf}
If $\alpha\in (0,1/2]$, $p=1$, Assumption \ref{hyp:setup} holds, then:
\begin{enumerate}
	\item If $\beta_0=b_0$, $\lim_{n\to\infty} P\left(T_n(b_0)>z_{1-\alpha}(p)| T_{1,n} \le q_{1,n},...,T_{J,n}\le q_{J,n}\right)\le \alpha$.
	\item $\lim_{n\to\infty} P\left(\beta_0\in(-\infty, \widehat{\beta} + \widehat{\Sigma}_{\beta}^{1/2} z_{1-\alpha}]| T_{1,n} \le q_{1,n},...,T_{J,n}\le q_{J,n}\right)\ge 1-\alpha$.
\end{enumerate}
\end{prop}
Similarly, if $q=1$, meaning that the pre-test estimator $\widehat{\theta}$ is of dimension one, one can apply Lemma \ref{lem:GCI} again, but now swapping the roles of $\widehat{\beta}$ and $\widehat{\theta}$, to show that conditional on not rejecting a unilateral pre-test, bilateral tests or CRs on $\beta_0$ are asymptotically conservative. Hence, while Assumption \ref{hyp:setup} does not apply to moment inequality models, the main take-away from Theorem \ref{thm:valid_post} still applies to moment inequality models with one inequality. On the other hand, unilateral tests or CRs on $\beta_0$ are not guaranteed to be conservative conditional on not rejecting a unilateral pre-test.\footnote{When $p=q=1$, the distortion does not seem large. For instance, if we consider a 5\% level test of $\theta_0\le 0$, we find by numerical evaluation that across all possible values of $\Sigma_{12}$, the conditional coverage of the confidence interval  $(-\infty,\widehat{\beta}+1.64\widehat{\sigma}_\beta]$ is (asymptotically) at least equal to 94.7\%.}


\subsection{Testing infinitely many constraints} 
\label{sub:extension_to_testing_infinitely_many_constraints}

The pre-tests in Assumption \ref{hyp:setup} are based on a finite-dimensional estimand $\theta_0$. However, as the next example shows, some pre-tests are in fact based on infinite-dimensional estimands.
\begin{exi}{average marginal effect: set-up.}
\label{ex:lin}
We observe a variable $Y$ and a continuously distributed variable $D$ with support $\mathcal{D}$. For any $d\in \mathcal{D}$, let $m(d)=\partial E(Y|D=d)/\partial d$. Our target parameter is
$$\beta_0:=E[m(D)],$$
the average marginal effect of $D$ on $Y$. To estimate $\beta_0$, we use $\widehat{\beta}$, the coefficient on $D$ in an OLS regression of $Y$ on $(1,D)'$.
The null of valid specification is $E[Y|D]=\alpha_0+\beta_0D$, where $\alpha_0:=E[Y-\beta_0D]$. This is equivalent to the following, infinite-dimensional restriction:
\begin{equation}\label{eq:testable_implication_OLS}
\zeta_0(d):=E[(Y-\alpha_0-\beta_0 D)\ind{D\le d}]=0, \quad\forall d \in \mathcal{D}.
\end{equation}
\end{exi}

\medskip
To cover pre-tests based on infinite-dimensional estimands,  we introduce the following assumption.
\begin{hyp}~
	\begin{enumerate}
		\item \label{hyp:iid} We observe a sample $(U_i)_{i=1,...,n}$ of i.i.d. random vectors.
		\item \label{hyp:Pf} $\theta_0$ satisfies $\theta_0(f):=E[f(U_1)]=0$, for all $f\in \F$, where $\mathcal{F}$ is a Donsker class of functions.\footnote{For the definition of a Donsker class, see, e.g., \cite{VdV_Wellner}, p.130.} Similarly, $\widehat{\theta}(f)=\frac{1}{n}\sum_{i=1}^n f(U_i)$. Also,
		\begin{equation}
	\sqrt{n}\left(\widehat{\beta}-\beta_0\right) = n^{-1/2}\sum_{i=1}^n \psi(U_i) +o_P(1),		
			\label{eq:as_lin}
		\end{equation}
where $E[\|\psi(U_1)\|^2]<\infty$, $E[\psi(U_1)]=0$ and $V(\psi(U_1))$ is nonsingular.
		\item \label{hyp:test} We consider $J$ pre-tests of $H_0: \theta_0=0$ based on the statistics $(T_{1,n},...,T_{J,n}) \in\R^J$ that satisfy, under $H_0$,
		\begin{equation}
		T_{j,n} = T_j(n^{1/2}\widehat{\theta})+o_P(1).
			\label{eq:Tn}
		\end{equation}
		Moreover, $T_j$ is convex, $T_j(-\theta)=T_j(\theta)\ge 0$ and $T_j(0)=0$.\footnote{Formally, $T_j$ is defined on $\ell^\infty(\F)$, the set of bounded functions from $\F$ to $\R$, and in \eqref{eq:Tn}, $n^{1/2}\widehat{\theta}$ is seen as an element of $\ell^\infty(\F)$.} Also, $T_j$ is continuous with respect to the uniform norm $\|\theta\|_{\F}:=\sup_{f\in\F} |\theta(f)|$. The critical region of the $j$-th test is $\{T_{j,n}> q_{j,n}\}$, where $q_{j,n}$ is a random variable satisfying $q_{j,n}\convP q_j>0$.
	\end{enumerate}
	\label{hyp:setup_process}
\end{hyp}
Contrary to Assumption \ref{hyp:setup}, Assumption \ref{hyp:setup_process} allows for pre-tests that rely on infinite-dimensional estimands: the class $\F$ in Assumption \ref{hyp:setup_process}.\ref{hyp:Pf} may be infinite and even uncountable. Then, the continuity condition in Assumption \ref{hyp:setup_process}.\ref{hyp:test} is important, to apply the continuous mapping theorem. Note also that Assumption \ref{hyp:setup_process} does not allow for estimators converging at rates lower than $n^{1/2}$ or for non-independent variables. However, we conjecture that Theorem \ref{thm:valid_post_infinite} below still holds if we assume, instead of Assumption \ref{hyp:setup_process}.\ref{hyp:iid} and \ref{hyp:setup_process}.\ref{hyp:Pf}, the weak convergence of $(\widehat{\Sigma}_\beta^{-1/2}(\widehat{\beta} -\beta_0),\widehat{\Sigma}_\theta^{-1/2}(\widehat{\theta} -\theta_0))$ to a Gaussian process.

\begin{exi}{average marginal effect: pre-test.}
Condition \eqref{eq:testable_implication_OLS} can be tested using the Kol-mogorov-Smirnov or Cram\'er-von Mises tests of linearity proposed in \cite{stute1997nonparametric} and \cite{stute1998bootstrap}.\footnote{\label{foot:Horowitz} Contrary to, e.g., the test of \cite{horowitz2001adaptive}, this test has not been shown to be (and may not be) minimax-optimal. However, it has the advantage of being free of tuning parameters. Moreover, the test of \cite{horowitz2001adaptive} does not satisfy Assumption \ref{hyp:setup_process}.}
Assume that we observe a sample of $n$ i.i.d. vectors $U_i:=(D_i,Y_{i})'$, with $E[Y^4]<\infty$, $E[D^4]<\infty$ and $V(D)>0$.
Let $\gamma_0=(\alpha_0,\beta_0)'$ and $\widehat{\gamma}=(\widehat{\alpha},\widehat{\beta})$ be the OLS estimator in the regression of $Y$ on $(1,D)'$. Consider a Kolmogorov-Smirnov test statistic  (the same reasoning applies to a Cram\'er-von Mises statistic): $T_{1,n}= n^{1/2} \sup_{d\in \mathcal{D}}|\widehat{\zeta}(d)|$, with
$$\widehat{\zeta}(d):=\frac{1}{n}\sum_{i=1}^n \ind{D_i\le d}(Y_i - (\widehat{\alpha}+\widehat{\beta}D_i)).$$
By a uniform law of large numbers on $\sum_{i=1} \ind{D_i\le d} (1,D_i)'/n$ and standard results on regressions, we obtain
\begin{align}
\sqrt{n}\left(\widehat{\zeta}(d)-\zeta_0(d)\right) =&\frac{1}{\sqrt{n}}\sum_{i=1}^n \left[\ind{D_i\le d} - E[\ind{D\le d}(1,D)]E[(1,D)'(1,D)]^{-1}(1,D_i)'\right] \nonumber \\
& \; \times (Y_i - (1,D_i)\gamma_0) + o_P(1),\label{eq:approx_infinite_dim}
\end{align}
where the $o_P(1)$ is uniform over $d\in\mathcal{D}$. Let us define
\begin{align*}
\mathcal{F}:=& \bigg\{f_{d'}: (y,d)\mapsto \left[\ind{d\le d'}- E[\ind{D\le d'}(1,D)]E[(1,D)'(1,D)]^{-1}(1,d)'\right] \\
& \;\,\times (y - (1,d)\gamma_0),
~d'\in\mathcal{D}\bigg\}.	
\end{align*}
For all $f_d\in \mathcal{F}$, let $\theta_0(f_d):=E(f_d(U))$ and $\widehat{\theta}(f_d):=\frac{1}{n}\sum_{i=1}^n f_d(U_i)$. Then, if $E[Y|D]=\alpha_0 + \beta_0 D$,  we have $\theta_0(f_d)=\zeta_0(d)=0$ for all $d\in\mathcal{D}$. Also, standard results imply that $\mathcal{F}$ is Donsker and by \eqref{eq:approx_infinite_dim},
$$n^{1/2}\widehat{\zeta}(d)=n^{1/2}\widehat{\theta}(f_d)+o_P(1).$$
Moreover, by continuity of the supremum functional and the previous display, $T_{1,n}=T_1\left(n^{1/2}\widehat{\theta}\right)+o_P(1)$, with $T_1(\theta)=\sup_{f\in\mathcal{F}} |\theta(f)|$. Hence, Assumptions \ref{hyp:setup_process}.\ref{hyp:Pf} and \ref{hyp:setup_process}.\ref{hyp:test} hold.
\end{exi}

Beyond the previous example, Assumption \ref{hyp:setup_process} applies to tests of functional form, where one wants to test if $E(Y|D)$ is equal to, say, a polynomial of known degree $K$ in $D$, or more generally to tests of conditional moment restrictions such as $E[g(U,\nu_0)|X]=0$.

\begin{thm}
    If Assumption \ref{hyp:setup_process} holds, for any centro-symmetric convex set $C$,
	\begin{equation}
	\lim_{n\to\infty} P\left[\sqrt{n}\big(\widehat{\beta}-\beta_0\big) \in C \big| T_{1,n} \le q_{1,n},...,T_{J,n}\le q_{J,n}\right] \ge  P(\widetilde{Z}_1\in C),	
		\label{eq:conv_prob_condit_A2}
	\end{equation}
where $\widetilde{Z}_1\sim\mathcal{N}(0,V(\psi))$.\footnote{Assumption \ref{hyp:setup_process} does not guarantee that the statistics $T_{j,n}$ are measurable, as the empirical process indexed by $\mathcal{F}$ may not be measurable. If not, the result holds replacing $P$ on the left-hand side of \eqref{eq:conv_prob_condit_A2} by the outer probability $P^*$ \citep[see, e.g.][p.6]{VdV_Wellner}.}
	\label{thm:valid_post_infinite}
\end{thm}
Theorem \ref{thm:valid_post_infinite} follows from the same steps as Theorem \ref{thm:valid_post}, but also from the approximation of the limit distribution of $T_j\left(n^{1/2}\widehat{\theta}\right)$ by a convex function of a finite-dimensional Gaussian measure. Though we do not state them, Theorem \ref{thm:valid_post_infinite} readily implies results equivalent to those in Corollary \ref{cor:CR_tests} if Assumption \ref{hyp:setup_process} holds.


\section{Results under the alternative}
\label{sec:alt}

In this section, we consider results when $(\theta_0,\eta_0)\ne 0$. This requires modifying our modeling framework. With a fixed distribution such that $\theta_0\ne 0$, the power of the pre-tests typically converges to one when $n\to \infty$. Then, pre-testing can never lead to wrongly accept the null. In practice however, pre-tests may have limited power and may thus fail to reject under the alternative. To allow for this possibility, we consider, as is standard in the literature, local alternatives. Specifically, $(\theta_{0n},\eta_{0n})$, now indexed by $n$ to emphasize that the DGP changes with $n$, is different from 0 but tends to 0 as $n$ tends to infinity. More precisely, we assume
\begin{equation}
\delta_n:=r_n\times(\theta_{0n},\eta_{0n})\to \delta=(\delta_\theta,\delta_\eta).
	\label{eq:conv_delta_n}
\end{equation}
The term $r_n$ corresponds to the rate at which the estimators converge under the null. In our examples above, $r_n=n^{1/2}$ except in RDDs where $r_n=n^{2/5}$. 
Equation \eqref{eq:conv_delta_n} states that the distance between the distribution of the data and the null shrinks to zero at the same rate at which the estimators converge. In this framework, the power of the pre-tests does not converge to one, because as the sample size increases, deviations from the null become closer to zero and harder to detect. Note that we impose not only $r_n\theta_{0n} \to \delta_\theta$ but also $r_n\eta_{0n}\to \delta_\eta$, even though this is not needed for the asymptotic power of the pre-test to be less than one. We do so because $r_n\eta_{0n}\to \infty$ often implies that $\sqrt{n}(\widehat{\beta}-\beta_{0n})$ tends to infinity in probability. If so, CC and UC both converge to zero.

\medskip
Assumption \ref{hyp:setup_alt} below is the analogue of Assumption \ref{hyp:setup} under local alternatives.
\begin{hyp}~
	\begin{enumerate}
		\item \label{hyp:beta_hat2_bis} The sequence of DGPs satisfies $\delta_n\to \delta$ and
		\begin{equation}
\left(\widehat{\Sigma}_\beta^{-1/2}\left(\widehat{\beta}-\beta_{0n}\right), \widehat{\Sigma}_\theta^{-1/2}\widehat{\theta} \right) \convN{\begin{pmatrix} \mu_1(\delta) \\ \mu_2(\delta)\end{pmatrix},\begin{pmatrix}I_p & \Sigma_{12} \\ \Sigma_{12}' & \Sigma_{22}\end{pmatrix}},			
			\label{eq:as_nor_bias}
		\end{equation}
where $\Sigma_{22}$ is positive definite and $\mu_1(.)$ and $\mu_2(.)$ are continuous at $\delta=0$ with $\mu_1(0)=0$ and $\mu_2(0)=0$.
		\item The same conditions as in Assumption \ref{hyp:setup}.\ref{hyp:test2} hold, with $\theta_0$ replaced by $\theta_{0n}$ and ``if $\theta_0=0$'' replaced by ``if $\theta_{0n}$ belongs to a neighborhood of 0''.
	\end{enumerate}
	\label{hyp:setup_alt}
\end{hyp}
Assumption \ref{hyp:setup_alt}.\ref{hyp:beta_hat2_bis} requires that if the distance between the distribution of the data and the null of valid specification converges to zero at the same rate at which the estimators converge, then $\widehat{\beta}-\beta_{0n}$ and $\widehat{\theta}$ are asymptotically normal, potentially with first-order biases. If $\delta=0$, then the asymptotic biases are equal to zero, as would happen under the null. On the other hand, $\widehat{\beta}$ may have an asymptotic bias even if the testable part of the null hypothesis holds (i.e. $\delta_\theta=0$). Allowing for this generality is important: for instance, in the DID example, parallel trends before $t_0$ may not necessarily imply parallel trends after.\footnote{Similarly, under Assumption \ref{hyp:setup_alt}, having $\delta_\eta=0$ is not sufficient to have $\mu_2(\delta)=0$. This generality may not be as important: in our four leading examples, $\delta_\eta=0$ does imply $\mu_2(\delta)=0$.} We now give conditions under which Assumption \ref{hyp:setup_alt}.\ref{hyp:beta_hat2_bis} holds in three of our four examples.\footnote{We conjecture that it should be possible to exhibit conditions under which  Assumption \ref{hyp:setup_alt}.\ref{hyp:beta_hat2_bis}  also holds in the RDD example.} Hereafter, we let $\Sigma_\beta$ and $\Sigma_\theta$ denote the probability limit of $r_n^2\widehat{\Sigma}_\beta$ and $r_n^2\widehat{\Sigma}_\theta$ (we show that these limits  exist in all our examples).

\begin{exi}{DID: conditions under which Assumption \ref{hyp:setup_alt}.\ref{hyp:beta_hat2_bis}  holds.}
Suppose that for all $t\in\{1,...,T\}$, $G_n=G$, $Y_{n,t}(1)=\widetilde{Y}_t(1)$ and
\begin{equation}
Y_{n,t}(0)=\widetilde{Y}_t(0) + G\left(\theta_{0n,t}\ind{t<t_0-1}+ \eta_{0n,t-t_0+1}\ind{t\ge t_0}\right),	
	\label{eq:Y_under_alt_DID}
\end{equation}
where the distribution of $(G,\widetilde{Y}_1(0),\widetilde{Y}_1(1),...,\widetilde{Y}_T(0),\widetilde{Y}_T(1))$ does not vary with $n$, admits a second-order moment and for all $t\in\{1,...,T\}$,
\begin{equation}
E(\widetilde{Y}_{t}(0)-\widetilde{Y}_{t_0-1}(0)|G=1)=E(\widetilde{Y}_{t}(0)-\widetilde{Y}_{t_0-1}(0)|G=0).	
\label{eq:Y_under_alt_DID2}
\end{equation}
This DGP depends on $n$ only through the vanishing violations of parallel trends in \eqref{eq:Y_under_alt_DID}. While more general DGPs could be considered, it is easy to check that Assumption \ref{hyp:setup_alt}.\ref{hyp:beta_hat2_bis} holds in this setup. Specifically, if $\delta_n\to \delta$, it follows from the law of large numbers and the central limit theorem that $\Sigma_\beta$ and $\Sigma_\theta$ exist and \eqref{eq:as_nor_bias} holds, with $\mu_1(\delta)=\Sigma_\beta^{-1/2}\delta_\eta$, $\mu_2(\delta)=\Sigma_\theta^{-1/2}\delta_\theta$ and $\Sigma_{22}=I_q$.
\end{exi}

\begin{exi}{arguably exogenous: conditions under which Assumption \ref{hyp:setup_alt}.\ref{hyp:beta_hat2_bis}  holds.} Similarly, suppose that $D_n=D$, $Y_n(1)=\widetilde{Y}(1)$ and
\begin{equation}
	Y_n(0) = \widetilde{Y}(0)+D\eta_{0n}, \; X_{b,n}=\widetilde{X}_b+D\theta_{0n},
	\label{eq:DGP_AE_alt}
\end{equation}	
where the distribution of $(D,\widetilde{Y}(0),\widetilde{Y}(1),\widetilde{X}_b)$ does not vary with $n$, admits a second-order moment and
\begin{equation}
E[\widetilde{Y}(0)|D=0]=E[\widetilde{Y}(0)|D=1], \; E[\widetilde{X}_b|D=0]=E[\widetilde{X}_b|D=1].
	\label{eq:DGP_AE_alt2}
\end{equation}
Then, if $\delta_n\to \delta$, $\Sigma_\beta$ and $\Sigma_\theta$ exist and  \eqref{eq:as_nor_bias} holds, with again $\mu_1(\delta)=\Sigma_\beta^{-1/2}\delta_\eta$,  $\mu_2(\delta)=\Sigma_\theta^{-1/2}\delta_\theta$ and $\Sigma_{22}=I_q$.
\end{exi}

\begin{exi}{GMM: conditions under which Assumption \ref{hyp:setup_alt}.\ref{hyp:beta_hat2_bis} holds.} \cite{newey1985generalized} considers a setup in which $U_n$ has a density $f(\cdot, c_n)$ with respect to a certain measure, for some vector $c_n$ satisfying $c_n=c_0 + \delta_c/\sqrt{n}$ and $E[g(U_n,\nu_0)]=0$ if $\delta_c=0$. As above, let $\widehat{\beta}:=\sum_{i=1}^n h(U_i,\widehat{\nu}_W)/n$ and $\widehat{\theta}:= W^{*1/2}\sum_{i=1}^ng(U_i,\widehat{\nu}_{W^*})/n$. Then, under Assumptions 1-7 in \cite{newey1985generalized}, his Lemma 1 together with a law of large numbers and Slutsky's theorem imply that (i) $\delta_n\to \delta=(\delta_\theta,0)$, with $\delta_\theta:=\int g(u,\nu_0)\deriv{f}{c}(u,c_0)\delta_cdu$ (recall that $\eta_{0n}=0$ here), and (ii) Eq. \eqref{eq:as_nor_bias} holds,  with $\mu_1(\delta)= -\Sigma_\beta^{-1/2}J_h(J_g'WJ_g)^{-1}J_g' W\delta_\theta$, $\mu_2(\delta) = \Sigma_\theta^{-1/2}\Pi W^{* \,1/2}\delta_\theta$ and $\Sigma_{22}=I_q$.
\end{exi}

\subsection{Results in a neighborhood of the null}

Theorem \ref{thm:altern} below shows that when the estimators used in the pre-tests $\widehat{\theta}$ and the estimators of interest $\widehat{\beta}$ are not asymptotically independent, and the pre-tests are either sup-$t$ tests or $F$-tests, then there exist neighborhoods around the null ($\delta=0$) such that if $\delta$ belongs to that neighborhood, CRs' conditional coverage (CC) is at least as large as their nominal coverage (NC) and their unconditional coverage (UC).
\begin{thm}\label{thm:altern}
	Suppose that Assumption \ref{hyp:setup_alt} holds with $J=1$ and either (i) $T_1(x)=\max(|x_1|,$ $...,|x_q|)$ and $p=1$ or (ii) $T_1(x)=x'x$ and $\Sigma_{22}=I_q$. Then:
\begin{enumerate}
\item If $\Sigma_{12}\ne 0$, there exists a neighborhood $\mathcal{V}$ of $0\in\R^{q+r}$ such that if $\delta\in \mathcal{V}$, then:
\begin{equation}
\lim_{n\to\infty} P\left(\beta_{0n}\in\CR| T_{1,n} \le q_{1,n}\right)\ge 1-\alpha.	
	\label{eq:valid_alt}
\end{equation}
\item There exists a neighborhood $\mathcal{V}'$ of $0\in\R^{q+r}$ such that if $\delta\in \mathcal{V}'$, then:
\begin{equation}
\lim_{n\to\infty} P\left(\beta_{0n}\in\CR| T_{1,n} \le q_{1,n}\right)\ge \lim_{n\to\infty} P\left(\beta_{0n}\in\CR\right).
	\label{eq:better_than_UC}
\end{equation}
\end{enumerate}
\end{thm}

The intuition of the proof of Theorem \ref{thm:altern} is as follows. First, by Proposition \ref{prop:conserv}, for sup-$t$ tests and $F$-tests, CRs' CC is strictly greater than their NC under the null if $\widehat{\theta}$ and $\widehat{\beta}$ are not asymptotically independent. Then, by continuity, there must be a neighborhood where CRs' CC is still weakly greater than their NC. By the same reasoning, there must also be a neighborhood around the null where CRs' CC is weakly greater than their UC, as UC is equal to NC under the null. Theorem \ref{thm:altern} focuses on CRs, but a similar result holds for tests.
\begin{exi}{DID: remark on Theorem \ref{thm:altern}.}
Even if $\theta_{0n}=0$ and $\eta_{0n}\ne 0$ (parallel trends holds before but not after treatment), it follows from Theorem \ref{thm:altern} that tests of differential pre-trends can offer a benefit, by ensuring a CC at least as large as NC and UC, in a neighborhood around the null.
\end{exi}

\begin{exi}{arguably exogenous: remark on Theorem \ref{thm:altern}.}
Similarly, if $\theta_{0n}=0$ and $\eta_{0n}\ne 0$, meaning that balance holds on pre-determined covariates but not on unobservables, it follows from Theorem \ref{thm:altern} that balancing tests on $X_b$ can offer a benefit, by ensuring a CC at least as large as NC and UC, in a neighborhood around the null.
\end{exi}

\begin{exi}{GMM: remark on Theorem \ref{thm:altern}.}
Inasmuch as one favors some form of protection against misspecification over valid inference under the null, Theorem \ref{thm:altern} can motivate the use of a non-optimal GMM estimator: if $h(U_i,\nu_0)$ does not depend on $U_i$ and one uses the optimal GMM estimator $\widehat{\beta}_{W^*}$, then $\Sigma_{12}=0$, so Point 1 of Theorem \ref{thm:altern} does not apply.
\end{exi}

\begin{figure}[H]
	\begin{center}
		\includegraphics[scale=0.85,trim=20mm 85mm 10mm 80mm, clip=true]{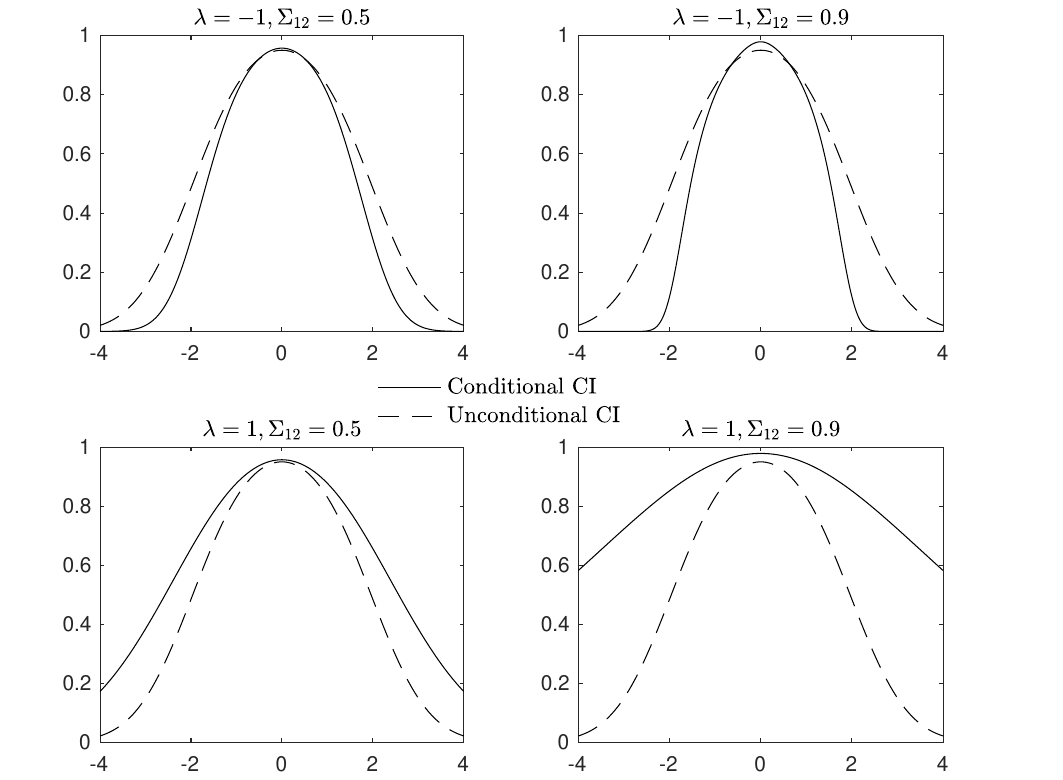}
	\end{center}
	{\footnotesize Notes: the x-axis corresponds to $\mu_2:=\mu_2(\delta)$. We consider $p=q=1$, with 	$\mu_1(\delta)=\lambda \mu_2$, $\Sigma_{22}=1$ and four different values of $(\lambda,\Sigma_{12})$.}
	\begin{center}
	\caption{Conditional and unconditional coverage rates of a 95\% level CI under local alternatives.}\label{fig:covg_local}
	\end{center}	
\end{figure}

\medskip
Theorem \ref{thm:altern} raises the following question: in practice, how large are the neighborhoods around the null where CRs' CC is not lower than their NC (resp. UC)? The answer to that question depends heavily on $(\mu_1(\delta),\mu_2(\delta),\Sigma_{11},\Sigma_{22},\Sigma_{12})$. Figure \ref{fig:covg_local} below considers the case where $p=q=1$: the pre-test is based on a scalar estimator, in which case the sup-$t$ test and the $F$-test are equivalent, and the parameter of interest is also scalar. Then, the figure shows the asymptotic CC and UC of $\CIU$, when $\mu_1(\delta)=\lambda \mu_2(\delta)$, $\Sigma_{11}=\Sigma_{22}=1$, as a function of $\mu_2:=\mu_2(\delta)$ and for four different values of $(\lambda,\Sigma_{12})$. $\mu_2$, the bias of the pre-test estimator divided by its standard deviation, can directly be mapped to the power of the pre-test. For instance, if $|\mu_2|=1.96$ (resp. $2.8$), the pre-test has a 50\% (resp. 80\%) chance of being rejected.

\medskip
In the top-left panel, $\Sigma_{12}=0.5$ and $\lambda=-1$: $\widehat{\beta}$ and $\widehat{\theta}$ are positively correlated, and the standardized bias of $\widehat{\beta}$ is the negative of that of $\widehat{\theta}$. In that DGP, under the null (i.e. $\mu_2=0$), CC of $\CIU$ is equal to 95.7\%, barely above 95\%. Accordingly, the neighborhood around $\mu_2=0$ where CC is above 95\% is small ($[-0.22,0.22]$), and the neighborhood where CC is above UC is also small ($[-0.51,0.51]$). At $|\mu_2|=1.96$, CC is equal to $33\%$, which is much lower than 95\%, and quite a bit lower than UC (50\%). At $|\mu_2|=2.8$, CC is equal to $4.2\%$, which is orders of magnitude lower than 95\%, and much below UC (20\%). In the top-right panel, $\Sigma_{12}=0.9$ and $\lambda=-1$. With very strongly positively correlated estimators, the neighborhoods where CC is above 95\% and above UC become larger ($[-0.36,0.36]$ and $[-0.63,0.63]$, respectively). On the other hand, CC declines more steeply when $|\mu_2|$ increases. For instance, at $|\mu_2|=1.96$, CC is equal to $14.4\%$. In the bottom-left panel, $\Sigma_{12}=0.5$ and $\lambda=1$: the standardized bias of $\widehat{\beta}$ is equal to that of $\widehat{\theta}$. Then, while the neighborhood where CC is above 95\% remains small ($[-0.29,0.29]$), CC is larger than UC for all $\mu_2$. The bottom-right panel shows that results are similar when $\Sigma_{12}=0.9$, though the neighborhood where CC is above 95\% becomes larger ($[-0.91,0.91]$).\footnote{Considering only positive values of $\Sigma_{12}$ is without loss of generality: if $\Sigma_{12}<0$ one can let $\widetilde{\theta}=-\widehat{\theta}$.}

\subsection{``Global'' result}\label{sub:unif}

The bottom panels of Figure \ref{fig:covg_local} suggest that under some conditions, CC is larger than UC for all $\delta$. The following theorem confirms this idea.

\begin{thm}\label{thm:altern_glob}
	Suppose that Assumption \ref{hyp:setup_alt} holds with $J=1$, $p=1$, $T_1(x)=x'x$ and $\Sigma_{22}=I_q$. If also $\mu_1(\delta)=\Sigma_{12} \mu_2(\delta)$ for all $\delta\in\R^{q+r}$, then \eqref{eq:better_than_UC} holds for all $\delta\in\R^{q+r}$.
\end{thm}

Our result relies on the same assumptions as Theorem \ref{thm:altern}, and three additional conditions. First, we need to have $p=1$: the target parameter should be a scalar. Second, we should have $\Sigma_{22}=I_q$. This essentially requires that once multiplied by the convergence rate squared, $\widehat{\Sigma}_\theta$ is consistent for the estimators' asymptotic variances under the null and under local alternatives, a condition that often holds in practice. Third, we need to have \(\mu_1(\delta)=\Sigma_{12}\mu_2(\delta)\): the standardized bias of \(\widehat\beta\) should equal the sum of the standardized biases of the components of \(\widehat\theta\), weighted by their correlations with \(\widehat\beta\). Under those additional conditions, \eqref{eq:better_than_UC} holds uniformly: CRs' CC is always greater than their UC.

\medskip
Theorem \ref{thm:altern_glob} only applies to $F$-tests. Its proof shares some similarities with that of Proposition \ref{prop:conserv}, but with important differences. Let $(Z_1(\delta),Z_2(\delta))$ be such that
$$\left[\widehat{\Sigma}_\beta^{-1/2}(\widehat{\beta}-\beta_{0n}), \widehat{\Sigma}_\theta^{-1/2}\widehat{\theta}\right]\convL (Z_1(\delta),Z_2(\delta)).$$
We cannot apply Anderson's theorem to $Z_1(\delta)|Z_2(\delta)'Z_2(\delta)$, as in Proposition \ref{prop:conserv}, because its distribution is not centered. Instead, we apply it to $N(\delta):=Z_1(\delta) - \Sigma_{12}Z_2(\delta)$: when $\mu_1(\delta)=\Sigma_{12}\mu_2(\delta)$, $N(\delta)$ has mean zero. We also exploit $\mu_1(\delta)=\Sigma_{12}\mu_2(\delta)$ to establish asymptotic independence between $Z_1(\delta)^2$ and $Z_2(\delta)'Z_2(\delta)$ conditional on $V:=[\Sigma_{12} Z_2(\delta)]^2$.

\medskip
We now exhibit assumptions under which the additional conditions underlying Theorem \ref{thm:altern_glob} hold in the arguably exogenous example, before exhibiting assumptions under which $\mu_1(\delta)=\Sigma_{12}\mu_2(\delta)$ cannot hold in the DID example.
\begin{exi}{arguably exogenous: application of Theorem \ref{thm:altern_glob}.}
For any random variable $A$ and any random vector $B$, let $L(A|B)$ denote the linear regression of $A$ on $(1,B)$. Suppose, in addition to \eqref{eq:DGP_AE_alt}-\eqref{eq:DGP_AE_alt2}, that
\begin{equation}\label{eq:linear_model}
L[Y_n(0)|X_{b,n},D,X_{b,n}D]=\alpha_{0}+\alpha_{1}'X_{b,n}.
\end{equation}
This states that when regressing $Y_n(0)$ on $(1,X_{b,n},D,X_{b,n}D)$, the coefficients on $D$ and $X_{b,n}D$ are equal to zero: $D$ is exogenous once the variables in the balancing tests are controlled for. Suppose also that treatment effects are uncorrelated with $X_{b,n}$ among the treated:
\begin{equation}\label{eq:nocov_effects_X}
\Cov(Y_n(1)-Y_n(0),X_{b,n}|D=1)=0,
\end{equation}
Then, we show in Appendix \ref{ssub:verification_of_theorem_alt_glob} that the conditions of Theorem \ref{thm:altern_glob} hold.
\end{exi}

\smallskip
\begin{counterexi}{DID with differential monotone trends and AR(1) errors.}
Consider the following particular case of \eqref{eq:Y_under_alt_DID}-\eqref{eq:Y_under_alt_DID2}:
\begin{align}
Y_{i,t,n}(0)&=\alpha_{i}+\gamma_{t} + \ind{G_{i,n}=1}\lambda_{n,t}+\eps_{i,t},\quad \eps_{i,t}=\rho \eps_{i,t-1}+u_{i,t},\label{eq:monot_trends}\\
\lambda_{n,t_0-1}=0 & \text{ and either }  \lambda_{n,1}<...<\lambda_{n,T} \text{ or }  \lambda_{n,1}>...>\lambda_{n,T}, \label{eq:monot_trends2}
\end{align}
where the $(\gamma_t,\lambda_{n,t})_{t\ge 1}$ are non-stochastic, $u_{i,t}$ is independent of $G_i$, mean zero, and i.i.d. across $i$ and $t$, and where $|\rho|< 1$. Condition \eqref{eq:monot_trends2} holds in particular with differential linear trends, for which $\lambda_{n,t}=\lambda_n(t-t_0+1)$. We focus here on $\beta_{0n}:=E[Y_{t_0-1+\ell,n}(1)-Y_{t_0-1+\ell,n}(0)|G=1]$ and assume for simplicity that treatment effects are constant ($Y_{i,t_0-1+\ell,n}(1)-Y_{i,t_0-1+\ell,n}(0)=\beta_{0n}$). Under these assumptions, we show in Appendix \ref{sub:details_on_example_did} that $\mu_1(\delta)\ne \Sigma_{12}\mu_2(\delta)$ for some $\delta$. Thus, we cannot apply Theorem \ref{thm:altern_glob}. When $T=t_0=3$, meaning that there is only one pre-trend and one event-study estimator, the intuition for the result is very simple: $\Sigma_{12}$ is strictly positive (this follows from the formula for $C_{\ell,t}$ below Proposition \ref{prop:conserv}), whereas $\mu_1(\delta)$ and $\mu_2(\delta)$ are of opposite signs with differential monotone trends.
\end{counterexi}

Theorem \ref{thm:altern_glob} gives a simple sufficient condition, $\mu_1(\delta)=\Sigma_{12}\mu_2(\delta)$, under which CC is larger than UC for arbitrarily large violations of the null. Importantly, this conclusion is not confined to DGPs satisfying that equality exactly. In Appendix \ref{app:num}, we show that, when $p=q=1$, CC can remain larger than UC under substantial departures from it. This is particularly relevant for the arguably-exogenous example. There, $\mu_1(\delta)=\Sigma_{12}\mu_2(\delta)$ essentially requires that all of the bias in $\widehat{\beta}$ is explained by imbalances in the covariates included in the balancing test. In practice, some bias may instead come from omitted variables that are not included in those tests. Our numerical results allow this omitted-variable component to be quite large: for instance, CC remains larger than UC when the omitted-variable bias is as large as the bias attributable to the observed covariate imbalance, over a broad range of $\delta$.

\subsection{Comparing CC \& UC, in DGPs calibrated to DID papers}\label{sub:calibration}

\paragraph{Motivation and set-up.}
In this section, we compare CC and UC of $\CIU$, in DGPs calibrated to the 12 DID studies in the meta-analysis conducted by \cite{roth2022pretest}.\footnote{Those are all the event-study DID papers published in the AER, AEJ:Policy, and AEJ:Applied between 2014 and 2018, with publicly available datasets.} We choose DID rather than cross-sectional examples (e.g. RCT, IV, matching, or RDD), because in the DID case $\Sigma_{12}$ is likely to be positive (this holds for ten of the 12 studies we consider, see Column (2) of Table \ref{tab:details_center} below), while  $\mu_1(\delta)$ and $\mu_2(\delta)$ are likely to be of opposite signs (this for instance holds with differential monotone trends). Then, Theorem \ref{thm:altern_glob} and the numerical result in Appendix \ref{app:num} are unlikely to apply, thus making DID more adversarial than these other examples for the comparison of CC and UC. For each paper, we consider the event-study graph revisited by \cite{roth2022pretest}, and restrict attention to the first pre-trend estimate and to the first event-study estimate, so that $p=q=1$. Then, we calibrate $\mu_2(\delta)$ to the t-statistic of the pre-trend and $\Sigma_{12}$ to the correlation between the pre-trend and the event-study estimator. Finally, we let $\mu_1(\delta)=-(\Sigma_{\theta}/\Sigma_{\beta})^{1/2}\mu_2(\delta)$, which holds if we have differential linear trends. We compute CC and UC for these 12 values of $(\mu_1(\delta),\mu_2(\delta),\Sigma_{12})$, for $\alpha=0.05$. Note that setting $\mu_2(\delta)$ at the realized t-stat of the pre-trend is an adversarial choice: even if trends were perfectly parallel in the twelve papers, the twelve $\mu_2(\delta)$ would correspond to realizations of twelve standard normals, with some of them being mechanically large, thus leading to a low CC and UC.

\paragraph{Results.}
Results are shown in Columns (1) to (5) of Table \ref{tab:details_center} below. The pre-trend is significant at the 5\% level in only one paper. Yet, violations of parallel trends calibrated to mostly insignificant pre-trend estimates can lead to significant size distortions: on average, UC and CC are respectively equal to 81.0\%
and 79.2\%, far below the 95\% NC of $\CIU$. At the same time, there is only one value of $(\mu_1(\delta),\mu_2(\delta),\Sigma_{12})$ for which CC is more than 10 percentage points smaller than UC. Among the remaining values, there are four for which CC is larger than UC and three for which CC is less than 1 percentage point smaller than UC. Thus, for those 12 values of $(\mu_1(\delta),\mu_2(\delta),\Sigma_{12})$, pre-testing rarely decreases much the coverage of $\CIU$. In Table \ref{tab:details_extremes} in the appendix, we keep the same values of $\Sigma_{12}$, but we calibrate $\mu_2(\delta)$ to the lower and upper bounds of the 95\% CI of the first pre-trend estimate, while still letting $\mu_1(\delta)=-(\Sigma_{\theta}/\Sigma_{\beta})^{1/2}\mu_2(\delta)$. Unsurprisingly, larger differential trends worsen the coverage  of $\CIU$: across these 24 values of $(\mu_1(\delta),\mu_2(\delta),\Sigma_{12})$, UC and CC are respectively equal to 53.8\% and 49.7\% on average. Now, there are two values of $(\mu_1(\delta),\mu_2(\delta),\Sigma_{12})$ for which CC is more than 10 percentage points smaller than UC, but for those UC is lower than 70\%, thus implying that $\CIU$ is unreliable even without pre-testing.

\input{tables/tab_details_center}

\section{Concluding remarks}\label{sec:conclu}

Specification tests are often used to decide whether an identifying assumption is sufficiently credible for estimation and inference to proceed. A common concern is that conditioning on passing such a test may itself invalidate subsequent inference. Our results qualify this concern. Under the null that the identifying assumption holds, confidence intervals remain valid conditional on passing the specification test, regardless of the dependence between the test statistic and the estimator. In fact, their conditional coverage is weakly higher than unconditional coverage. When the identifying assumption is violated, conditioning on passing the test can reduce coverage, but under a condition that can hold in some leading examples it instead increases coverage relative to not pre-testing. Of course, these results do not alleviate all concerns with specification tests. In particular, they may have low power, so passing a test need not provide strong evidence that the identifying assumption holds.

\medskip
An important limitation is that we study a researcher who considers one identifying assumption and reports an estimate only if the corresponding specification test is passed. In practice, researchers may respond to a rejection by modifying their specification and testing again until they obtain one that passes. For example, after rejecting a pre-trend test in a DID design, a researcher might add control variables and test again. Our results do not cover such sequential specification searches. Understanding inference after repeated testing and specification adjustment is an important direction for future research.

\medskip
Finally, our results also have perhaps unexpected implications for practitioners. For instance, inasmuch as one favors some form of protection against misspecification over valid inference under the null,  one may prefer using a non-optimal rather than optimal GMM estimator. Also,  our ``global'' and numerical results under local alternatives highlight that pre-testing may be more contentious in some contexts (e.g., DIDs) than in others (e.g., randomized experiments, IV studies). As these results require specific relationships between the two asymptotic biases and the asymptotic correlation of the test statistic and the estimator, they also imply that some specification tests may be more desirable than others. Again, we leave for future research the question of how to best choose the specification test from this perspective.

\newpage
\bibliography{biblio}

\appendix

\section{Numerical results under the alternative}\label{app:num} 

Theorem \ref{thm:altern} above shows that under weak conditions, CC of $\CI$ is larger than its UC in a neighborhood of the null, but the result is not informative on the neighborhood itself. Theorem \ref{thm:altern_glob} shows that under stronger conditions, notably $\mu_1(\delta)=\Sigma_{12}\mu_2(\delta)$, the neighborhood is actually the full space. Point 1 of the following numerical result gives quantitative guarantees on the neighborhood when $p=q=1$, under a weaker restriction than $\mu_1(\delta)=\Sigma_{12}\mu_2(\delta)$. Point 2 provides guarantees on CC itself, when $\mu_1(\delta)=\Sigma_{12}\mu_2(\delta)$.

\begin{num}\label{num:altern_almost_glob}
Suppose that Assumption \ref{hyp:setup_alt} holds with $J=1$, $p=q=1$, $\Sigma_{11}=1$, $\Sigma_{22}=1$, $\Sigma_{12}=\Sigma_{12}$, $T_1(x)=x^2$, $\alpha =0.05$ and $q_1$ equal to the 95\% quantile of a chi-squared distribution with $1$ degree of freedom. Finally, let  $\zeta=10^{-5}$.
\begin{enumerate}
\item If $\mu_1(\delta)=(1+R(\delta))\Sigma_{12}\mu_2(\delta)$, then \eqref{eq:better_than_UC} holds:
    \begin{enumerate}
    \item for all $(R(\delta),\mu_2(\delta),\Sigma_{12})\in [0,1]\times [-3.9,3.9]\times [-1+\zeta,1-\zeta]$;
        \item for all $(R(\delta),\mu_2(\delta),\Sigma_{12})\in [-1,1]\times [-1.9,1.9]\times [-1+\zeta,1-\zeta]$;
        \item for all $(R(\delta),\mu_2(\delta),\Sigma_{12})\in [-0.85,1]\times [-2.8,2.8]\times [-1+\zeta,1-\zeta]$.
    \end{enumerate}
\item If $\mu_1(\delta)=\Sigma_{12}\mu_2(\delta)$, then:
\begin{enumerate}
\item for all $|\mu_2(\delta)| \le 0.84$, $\lim_{n\to\infty} P\left(\beta_{0n}\in\CI| T_{1,n} \le q_{1,n}\right) \ge 0.95$;
\item for all $|\mu_2(\delta)| \le 2.8$, $P\lim_{n\to\infty} \left(\beta_{0n}\in\CI| T_{1,n} \le q_{1,n}\right)\ge 0.88.$
\end{enumerate}
\end{enumerate}
\end{num}

To establish, say, the first result in Point 1, we minimize $P(\beta_{0n}\in\CIU| T_{1,n} \le q_{1,n})/P(\beta_{0n}\in$ $\CIU)$ across all $(R(\delta),\mu_2(\delta),\Sigma_{12})\in [0,1]\times [-3.9,3.9]\times [-1+\zeta,1-\zeta]$, using Matlab's \st{fminunc} function. For 500 randomly chosen starting values, the objective function is always greater than or equal to one at the minimum, so the result seems robust to the fact that the objective function exhibits local minima. We proceed similarly to establish all other numerical results.

\medskip
Point 1.a) means that for essentially any value of $(\mu_2(\delta),\Sigma_{12})$,\footnote{The result is probably true for any value of $(\mu_2(\delta),\Sigma_{12})$, though up to Matlab's numerical precision we can only establish it numerically for $(\mu_2(\delta),\Sigma_{12})\in [-3.9,3.9]\times [-1+\zeta,1-\zeta]$.} CC is larger than UC if the standardized bias of $\widehat{\beta}$ belongs to $\Sigma_{12}\mu_2(\delta)\times[1,2]$.\footnote{If $\Sigma_{12}\mu_2(\delta)<0$, this interval should be understood as $[2\Sigma_{12}\mu_2(\delta),\Sigma_{12}\mu_2(\delta)]$.} Point 1.b) means that for essentially any value of $\Sigma_{12}$ and for small to moderately large values of $\mu_2(\delta)$, CC is still larger than UC if the standardized bias of $\widehat{\beta}$ belongs to $\Sigma_{12}\mu_2(\delta)\times[0,2]$. Finally, Point 1.c) means that for essentially any value of $\Sigma_{12}$ and any value of $\mu_2(\delta)$ for which the pre-test has less than 80\% chances of being rejected, CC is still larger than UC if the standardized bias of $\widehat{\beta}$ belongs to $\Sigma_{12}\mu_2(\delta)\times[0.15,2]$. Thus, \eqref{eq:better_than_UC} can still hold under fairly large departures from $\mu_1(\delta)=\Sigma_{12}\mu_2(\delta)$.

\medskip
While \eqref{eq:better_than_UC} holds uniformly when $\mu_1(\delta)=\Sigma_{12}\mu_2(\delta)$, \eqref{eq:valid_alt} does not: $\mu_1(\delta)=\Sigma_{12}\mu_2(\delta)$ is not sufficient to ensure that the CI's CC is least as large as its NC. Still, Point 2 of Numerical Result \ref{num:altern_almost_glob} shows that if $p=q=1$ and $\alpha=0.05$, \eqref{eq:valid_alt} holds for all $|\mu_2(\delta)| \le 0.84$. Moreover, for all $|\mu_2(\delta)| \le 2.8$, the coverage rate of the 95\%-level CI for $\beta_{0n}$ cannot be lower than 88\%. By contrast, UC can be as low as 20\% in such cases.

\medskip
Point 1 of Numerical Result \ref{num:altern_almost_glob} can be fruitfully applied to the arguably exogenous example. There, rationalizing $\mu_1(\delta)=\Sigma_{12}\mu_2(\delta)$ requires in particular assuming that the treatment is uncorrelated to $Y_n(0)$ once the covariates in the balancing tests are controlled for. This is restrictive: controlling for $X_{b,n}$ may not be enough to ensure that the treatment is exogenous. We now allow for effects of omitted variables.

\begin{exi}{arguably exogenous: application of Point 1 of Numerical Result \ref{num:altern_almost_glob}.}
Point 1 of Numerical Result \ref{num:altern_almost_glob} requires that $q=1$. With an $F$-test based on $\widehat{\theta}$, this fails whenever $X_{b,n}$ is not scalar. Accordingly, assume one instead uses a t-test based on $\widehat{\alpha}_1'\widehat{\theta}$, where $\widehat{\alpha}_1$ is the coefficient of $X_{b,n}$ from a regression of $Y_{i,n}$ on $(1,X_{b,i,n}')$ in the untreated sample.\footnote{For this test, we can either consider a variance estimator accounting for variability of $\widehat{\alpha}_1$ or not: in our setup this will not matter asymptotically.} The rationale for that test is to assess if imbalances in $X_{b,n}$ can meaningfully bias $\widehat{\beta}$. Then, assume that \eqref{eq:DGP_AE_alt}-\eqref{eq:DGP_AE_alt2} and \eqref{eq:nocov_effects_X} hold, but \eqref{eq:linear_model} does not. Instead, suppose that
\begin{equation}\label{eq:linear_model_OVB}
L[Y_n(0)|X_{b,n},D_n,X_{b,n}D_n,W_n]=\alpha_{0}+\alpha_{1}'X_{b,n} + W_n,
\end{equation}
with $\alpha_1\ne 0$. Here $W_n$ is a (potentially unobserved) variable not included in the balancing tests. \eqref{eq:linear_model_OVB} requires that in the regression of $Y_n(0)$ on $(1,X_{b,n},D_n,X_{b,n}D_n,W_n)$, the coefficients on $D_n$ and $X_{b,n}D_n$ are equal to zero: the treatment is exogenous once $X_{b,n}$ and the omitted variable are controlled for.\footnote{Assuming that $W_n$ is scalar is without loss of generality: with a vector of omitted variables $O_n$, if $L[Y_n(0)|X_{b,n},D_n,X_{b,n}D_n, O_n]=\alpha_{0}+\alpha_{1}'X_{b,n} + \alpha_{2}'O_n$, then \eqref{eq:linear_model_OVB} holds with $W_n=\alpha_{2}'O_n$.} We also assume that $W_n=\widetilde{W}+D\tau_n$, where $\Cov(\widetilde{W},D)=0$ and the distribution of $(\widetilde{Y}(1),\widetilde{Y}(0),\widetilde{X}_b,\widetilde{W})$ does not depend on $n$. By projecting $\widetilde{W}$ onto $\widetilde{X}_{b}$, we can assume without loss of generality that these variables are uncorrelated but we assume more here, namely that $\Cov(\widetilde{W},\widetilde{X}_{b}|D=d)=0$ for $d=0,1$. We prove in Appendix \ref{ssub:verification_of_numerical_result} that under these conditions, Point 1 of Numerical Result \ref{num:altern_almost_glob} holds, with  $R(\delta)=\delta_\tau/\alpha_1'\delta_\theta$, where $\delta_\tau:=\lim_{n\to \infty}n^{1/2}\tau_n$ (we also show that this limit exists and equals $\delta_\eta - \alpha_1'\delta_\theta$).

\medskip
Then, Point 1.a) of Numerical Result \ref{num:altern_almost_glob} implies that if $\delta_\tau$, the bias coming from variables not included in the balancing tests, is of the same sign as, and smaller in absolute value than, $\alpha_1'\delta_\theta$, the bias induced by $X_b$, then the CI's CC is essentially always larger than its UC. Similarly, Point 1.b) implies that if $|\delta_\tau|\leq |\alpha_1'\delta_\theta|$, then CC is larger than UC for moderately large values of $\alpha_1'\delta_\theta$. Finally, Point 1.c) implies that if $\delta_\tau$ is included between $-0.85$ and $1$ times $\alpha_1'\delta_\theta$, then CC is larger than UC for any value of $\alpha_1'\delta_\theta$ with less than 80\% chances of being detected.
\end{exi}


\section{Proofs}

\subsection{Theorems \ref{thm:valid_post} and \ref{thm:valid_post_infinite}} 

\subsubsection{Theorem \ref{thm:valid_post_infinite}} 
\label{ssub:under_assumption2}

We begin by the proof of Theorem \ref{thm:valid_post_infinite}, which is the more complicated one. Let us define $Z_n:=\sqrt{n}(\widehat{\beta}-\beta_0)$,  $T_n:=\max_{j=1,...,J} [T_{j,n} - q_{j,n}]$ and $T(\theta):=\max_{j=1,...,J} [T_j(\theta)-q_j]$. $T$ is convex and continuous, as the maximum of $J$ convex and continuous functionals. Moreover,
$$P(Z_n\in C,T_{1,n}\le q_{1,n},...,T_{J,n}\le q_{J,n}) = P(Z_n\in C, T_n \le 0).$$
Let $G_n:=n^{1/2}(P_n-P)$ denote the empirical process  indexed by $\mathcal{G}:=\F\cup \{\psi^1,...,\psi^d\}$. By Points \ref{hyp:Pf} and \ref{hyp:test} of Assumption \ref{hyp:setup_process},
\begin{equation}
(Z_n, T_n) = (G_n\psi, T(n^{1/2}(\widehat{\theta} - \theta)) + o_P(1).	
	\label{eq:approx}
\end{equation}
As union of Donsker classes, $\mathcal{G}$ is Donsker \citep[see][p.136]{VdV_Wellner}. Hence, $G_n$ is asymptotically Gaussian; we denote by $G$ its limit. By \eqref{eq:approx} and the continuous mapping theorem,
$$(Z_n, T_n) \convL (G\psi, T(G)).$$
Let $\partial A$ denote the boundary of the set $A$. The boundary of $C\times (-\infty,0]$ satisfies
\begin{equation}
\partial (C\times (-\infty,0]) = (\partial C \times (-\infty,0]) \cup (C \times \{0\}).	
	\label{eq:boundary}
\end{equation}
Moreover, since $C$ is convex, $\partial C$ has Lebesgue measure zero \citep[see, e.g.][p.90]{lang1986note}. Since $V(\psi(U_1))$ is nonsingular,  the distribution of $G\psi$ is absolutely continuous with respect to the Lebesgue measure. Hence,   $P(G\psi \in \partial C)=0$. Now, by Theorem 11.1 and Problem 11.3 in \cite{davydov1998local} and continuity of $T$, the cumulative distribution function $H$ of $T(G)$ is strictly increasing on $(r_0,\infty)$, with $r_0:=\inf_{\theta\in\ell^\infty(\mathcal{F})} T(\theta)=\max_{j=1,...,J} -q_j<0$. Hence, $P(T(G)=0)=0$ and $H(0)>0$. Then, in view of \eqref{eq:boundary},
$$P((G\psi, T(G))\in\partial (C\times (-\infty,0]))=0.$$
Thus, by Portmanteau's lemma \citep[see, e.g., Lemma 18.9 in][]{vandervaart00},
$$\lim_{n\to\infty} P^*(Z_n\in C, T_n \le 0) = P(G\psi \in C, T(G)\le 0),$$
where $P^*$ denotes the outer probability. Similarly, $\lim_{n\to\infty} P^*(T_n \le 0) = P(T(G)\le 0)=H(0)>0$. Hence,
\begin{equation}
\lim_{n\to\infty} P^*(Z_n\in C| T_n \le 0) = P(G\psi \in C| T(G)\le 0).	
	\label{eq:weak_conv}
\end{equation}
Now, fix $\eps,\delta>0$. Because $G$ is tight (as $\mathcal{F}$ is Donsker), there exists a compact set $\mathcal{K}\subset\ell^\infty(\mathcal{F})$ such that $P(G\in\mathcal{K})>1-\delta/2$. Since $T$ is continuous on $\ell^\infty(\mathcal{F})$ and $\mathcal{K}$ is compact, there exists $\eta>0$ such that for any $(G_1,G_2)\in\mathcal{K}\times \ell^\infty(\mathcal{F})$ such that $\|G_1-G_2\|_{\infty}<\eta$, $|T(G_1)-T(G_2)|<\eps$. Now, fix $\zeta>0$ and let $\rho_P(f):=P[(f -Pf)^2]^{1/2}$. Since $\F$ is totally bounded \citep[see][pp. 138-139]{VdV_Wellner}, there exist $K\ge 1$ and $(f_1,...,f_K)\in\F^K$ such that $\forall f\in\F$, $\min_{k=1,...,K} \rho_P(f,f_k)<\zeta \eta$. $G$ is also uniformly continuous with respect to $\rho_P$ in the sense that
$$\sup_{(f,g):\rho_P(f,g)<u} |G(f)-G(g)| \convP 0 \; \text{ as } u\to 0.$$
Let $\widetilde{G}(f)=G(\widetilde{f})$, where $\widetilde{f}=\arg\min_{g\in\{f_1,...,f_K\}} \rho_P(f-g)$. Then,
$$\|G-\widetilde{G}\|_\F = \sup_{f\in\mathcal{F}}|G(f-\widetilde{f})|\le \sup_{(f,g):\rho_P(f,g)<\zeta \eta} |G(f)-G(g)|.$$
By taking $\zeta$ small enough, we get that $\|G-\widetilde{G}\|_\F<\eta$ with probability larger than $1-\delta/2$. Then the probability of the event $\{G\in\mathcal{K}, \|G-\widetilde{G}\|_\F<\eta\}$ is larger than $1-\delta$. Moreover, under this event, $|T(G)-T(\widetilde{G})|<\eps$. Hence, using a union bound, we obtain
$$P(G\psi \in C,\, T(G)\le 0) \ge P(G\psi \in C,\, T(\widetilde{G})\le -\eps) -\delta,$$
As a result,
$$P(G\psi \in C\,|\, T(G)\le 0) \ge \frac{P(G\psi \in C,\; T(\widetilde{G})\le - \eps)-\delta}{H(0)}.$$
We can write $T(\widetilde{G})$ as $\widetilde{T}_K(Gf_1,...,Gf_K)$ for some convex function $\widetilde{T}_K$. Let $\mu$ denote the Gaussian distribution of $(G\psi, Gf_1,...,Gf_K)$. $\mu$ is a centered Gaussian distribution. Let $E=C \times \R^K$ and $F=\R^p \times \widetilde{T}_K^{-1}((-\infty, -\eps])$, with $q:=\min(q_1,...,q_J)$. Both $E$ and $F$ are convex. Moreover, they are centro-symmetric since  $T(\theta)=T(-\theta)$. Then,  by the Gaussian correlation inequality \citep{royen2014},
\begin{align*}
P(G\psi \in C,\; T(\widetilde{G})\le - \eps) = & \, \mu(E\cap F) \\
 \ge & \, \mu(E) \mu(F) \\
= & \, P(\widetilde{Z}_1 \in C)P(T(\widetilde{G})\le - \eps),
\end{align*}
where the last equality follows by definition of $\widetilde{Z}_1$. Using again the fact that $\|G-\widetilde{G}\|_\F<\eta$ with probability larger than $1-\delta$, we have $P(T(\widetilde{G})\le - \eps)\ge P(T(G)\le - 2\eps)-\delta$. Thus,
\begin{align*}
P(G\psi \in C\,|\, T(G)\le 0) \ge & \frac{P(\widetilde{Z}_1 \in C)(H(-2\eps) - \delta) - \delta}{H(0)}.
\end{align*}
Since $H$ is continuous at $0$ and $(\eps,\delta)$ were arbitrary, we finally obtain
$$P(G\psi \in C\,|\, T(G)\le 0) \ge P(\widetilde{Z}_1 \in C).$$
The result follows by combining the last display with \eqref{eq:weak_conv}.


\subsubsection{Theorem \ref{thm:valid_post}} 
\label{ssub:proof_under_assumption1}

The begining of the proof is almost the same as that of Theorem \ref{thm:valid_post_infinite}, so we just highlight the differences. First, since $T_j$ is a convex function on $\R^q$, it is continuous. Hence, $T(x):=\max_{j=1,...,J} [T_j(x)-q_j]$ is continuous (and convex). Then, reasoning as above, we obtain
\begin{equation}
\lim_{n\to\infty}P\left[\widehat{\Sigma}_\beta^{-1/2}\left(\widehat{\beta}-\beta_0\right) \in C \,\big|\,  T_n \le 0\right] = P(Z_1\in C \,|\, T(Z_2)\le 0),	
	\label{eq:conv_simple_case}
\end{equation}
where $T_n:=\max_{j=1,...,J} [T_{j,n} - q_{j,n}]$ and
\begin{equation}
\begin{pmatrix} Z_1 \\ Z_2 \end{pmatrix} \sim \mathcal{N}\left(0,\begin{pmatrix} I_p & \Sigma_{12} \\ \Sigma_{12}' & \Sigma_{22} \end{pmatrix}\right).	
	\label{eq:vect_gaussien}
\end{equation}
The result follows directly by the Gaussian correlation inequality.



\subsection{Proof of Corollary \ref{cor:CR_tests}} 
\label{sub:proof_of_corollary_ref_cor_cr_tests}

Point 2 follows directly from Point 1 so we focus on the latter. Suppose that Assumption \ref{hyp:setup} holds. Let $C=\{x\in\R^p:x'x\le q_{1-\alpha}(p)\}$. Then, if $\beta_0=b_0$,
$$F_n(b_0)\le q_{1-\alpha}(p)\Leftrightarrow \widehat{\Sigma}_\beta^{-1/2}\left(\widehat{\beta}-\beta_0\right) \in C.$$
Because $C$ is convex and centro-symmetric, we have, by Theorem \ref{thm:valid_post},
\begin{align*}
	& \lim_{n\to\infty} P\left[F_n(b_0)\le q_{1-\alpha}(p)\big| T_{1,n}\le q_{1,n},...,T_{J,n}\le q_{J,n}\right] \\
 = & \lim_{n\to\infty} P\left[ \widehat{\Sigma}_\beta^{-1/2}\left(\widehat{\beta}-\beta_0\right) \in C\, \big|\, T_{1,n}\le q_{1,n},...,T_{J,n}\le q_{J,n}\right] \\
 \ge & P(Z_1 \in C) = 1-\alpha.
\end{align*}
The result follows.


\subsection{Proof of Proposition \ref{prop:conserv}} 
\label{sub:proof_of_proposition_ref_prop_conserv}

The proof for Case (i) follows directly from \eqref{eq:conv_simple_case} with $p=1$ and Theorem 1 of \cite{royen2014}. Let us turn to Case (ii) and assume $\Sigma_{12}\ne 0$. By Equation \eqref{eq:conv_simple_case} and the form of $F_n(b_0)$, it suffices to prove that if $(Z_1,Z_2)$ satisfy \eqref{eq:vect_gaussien} with $\Sigma_{22}=I_q$ and $\Sigma_{12}\ne 0$, then, for all $y>0$ and any centro-symmetric, convex and compact set $C$ satisfying $P(Z_1\in C)>0$,
\begin{equation}
P(Z_1 \in C,F \le y) > P(Z_1\in C)\; P(F \le y),	
	\label{eq:ineg_for_conserv}
\end{equation}
with $F:=Z_2'Z_2$. First, remark that $N:=Z_1 - \Sigma_{12}Z_2\sim \mathcal{N}(0,I_p - \Sigma_{12}\Sigma_{12}')$  and is independent of $Z_2$. Moreover, $Z_2/F^{1/2}$, and thus $U:=-\Sigma_{12}Z_2/F^{1/2}$, are independent of $F$. Hence,
\begin{equation}
(Z_1,F)= (-F^{1/2}U+N,F),	
	\label{eq:repres}
\end{equation}
where $(F,N,U)$ are mutually independent. Next, we prove that $m(y):=P(Z_1 \in C|F= y)$ is non-increasing and not constant on $\R^+$. Let $\mu$ be the probability measure of $N$. By what precedes,
\begin{align}
m(y)= & P(-y^{1/2}U+N \in C) \notag \\
= & E\left[P(-y^{1/2}U+N \in C|U)\right] \notag \\
= & E\left[\mu(C + y^{1/2}U)\right], \label{eq:link_m_mu}
\end{align}
where the last equality follows since $U$ and $N$ are independent. Now, by Anderson's theorem \citep{anderson1955integral}, $y\mapsto \mu(C + y^{1/2}U)$, and in turn $m$, are non-increasing. Moreover,
\begin{equation}
P(U= 0)=P(\Sigma_{12}Z_2= 0)=0,	
	\label{eq:no_null_v}
\end{equation}
where the last equality follows because (i) $Z_2$ has a non-singular variance matrix and (ii) $\Sigma_{12}\ne 0$ implies that the kernel of $\Sigma_{12}$ has dimension $q'<q$. Now, because $C$ is compact, we have, for all $c\in C$, by the triangle inequality,
$$\|c+y^{1/2}U\|\ge y^{1/2}\|U\| - \|c\|\ge y^{1/2}\|U\| - M,$$
for some $M>0$. As a result,
$$\mu(C+y^{1/2} U) \le \mu\left(\left\{x\in\R^p:\,\|x\| \ge y^{1/2}\|U\| - M\right\}\right)\to 0 \; \text{almost surely (a.s.) as } y\to\infty.$$
Then, by \eqref{eq:link_m_mu} and the dominated convergence theorem,  $\lim_{y\to\infty}m(y)= 0$. Since $m(0)=\mu(C)>0$, $m$ is not constant on $\R^+$.

\medskip
Finally, by the law of iterated expectations, $$P(Z_1\in C,F \le y) = E[m(F)\ind{F\le y}].$$
Moreover, $m(\cdot)$ and $f\mapsto \ind{f\le y}$ are decreasing. By what precedes, $m(F)$ is not a.s. constant and neither is $\ind{F\le y}$, since $y>0$ and $F$ has support $\R^+$. Hence, by Theorems 1.1 and 1.2 of \cite{jakubowski2021complement}, we have
$$E[m(F)\ind{F\le y}] > E[m(F)]E[\ind{F\le y}].$$
Equation \eqref{eq:ineg_for_conserv} follows using the law of iterated expectations again.



\subsection{Proof of Lemma \ref{lem:GCI}} 
\label{sub:proof_of_lemma_ref_lem_gci}

Let $K_0=[-a,a]\times \R^{k-1}$, $K_-=(-\infty, -a]\times \R^{k-1}$ and $K_+=[a,\infty)\times \R^{k-1}$. By the Gaussian correlation inequality and symmetry of $\mu$,
\begin{align*}
	\mu(L)\left[\mu(K_0)+2\mu(K_+)\right] & = \mu(L) \\
	& = \mu(L \cap K_0) + \mu(L \cap K_-) + \mu(L \cap K_+) \\
	& \ge \mu(L) \mu(K_0) + 2\mu(L \cap K_+).
\end{align*}
Hence, $\mu(L)\mu(K_+)\ge \mu(L \cap K_+)$. As a result,
\begin{align*}
	\mu(L\cap K) & = \mu(L) - \mu(L \cap K_+) \\
	& \ge \mu(L) - \mu(L)\mu(K_+) \\
	& = \mu(L)\mu(K).
\end{align*}


\subsection{Proof of Proposition \ref{prop:unilat_inf}} 
\label{sub:proof_of_prop_unilat_inf}

By applying Lemma \ref{lem:GCI} instead of the initial Gaussian correlation inequality, we obtain the same result as Theorem \ref{thm:valid_post} but with $C=(-\infty,a]$. Then, the same reasoning as in Corollary \ref{cor:CR_tests} leads to the result.


\subsection{Proof of Theorem \ref{thm:altern}} 
\label{sub:proof_of_theorem_ref_thm_altern}

For the first point, Proposition \ref{prop:conserv} ensures that \eqref{eq:valid_alt} holds with strict inequality when $\delta=0$. Let $(Z_1(\delta),Z_2(\delta))$ be such that $(\widehat{\Sigma}_\beta^{-1/2}(\widehat{\beta}-\beta_{0n}),\widehat{\Sigma}_\theta^{-1/2}\widehat{\theta})\convL (Z_1(\delta),Z_2(\delta))$. Then, as in Theorem \ref{thm:valid_post}, $\lim_{n\to\infty} P\left(\beta_0\in\CR| T_{1,n} \le q_{1,n}\right)=P(Z_1(\delta)\in C|T(Z_2(\delta))\le 0)$ for some convex set $C$. By Assumption \ref{hyp:setup_alt} and the dominated convergence theorem, the right-hand side of the previous display is continuous in $\delta$. The result follows.

\medskip
Turning to the second point, note first that the result trivially holds when $\Sigma_{12}=0$. If not, by Proposition \ref{prop:conserv} again, \eqref{eq:better_than_UC} holds with strict inequality when $\delta=0$. The result follows by continuity of $\delta \mapsto \lim_{n\to\infty} P\left(\beta_0\in\CR| T_{1,n} \le q_{1,n}\right) - P(\beta_0\in\CR)$, which follows again by Assumption \ref{hyp:setup_alt} and the dominated convergence theorem.

\subsection{Proof of Theorem \ref{thm:altern_glob}}

The conditions ensure that  for any $\delta\in\R^{q+r}$,
$$\left(\widehat{\Sigma}_\beta^{-1/2}(\widehat{\beta}-\beta_0), \widehat{\Sigma}_\theta^{-1/2}\widehat{\theta}\right)\convL (Z_1(\delta),Z_2(\delta)),$$
where $(Z_1(\delta),Z_2(\delta))$ satisfy
$$(Z_1(\delta),Z_2(\delta))\sim \mathcal{N}\left(\begin{pmatrix} \Sigma_{12} \mu_2(\delta) \\ \mu_2(\delta)\end{pmatrix},\begin{pmatrix} 1 & \Sigma_{12} \\ \Sigma_{12}' & I_q \end{pmatrix}\right).$$	
To simplify the exposition, we omit the dependence in $\delta$ hereafter. Let $F:=Z_2'Z_2$, $V:=(\Sigma_{12}Z_2)^2$, and $\rho:=||\Sigma_{12}||$. As in Proposition \ref{prop:conserv}, it suffices to prove that
\begin{equation}
P(Z_1^2\le x, F\le y) \ge P(Z_1^2\le x)P(F\le y).	
	\label{eq:ineg_for_global}
\end{equation}
We proceed in four steps. First, we show that $Z_1^2\indep F|V$. Second, we prove that $g_x: v \mapsto P(Z_1^2\le x|V=v)$ is decreasing. Third, we show that $h_y: v \mapsto P(F\le y|V=v)$ is decreasing as well. The fourth step concludes. Note that we assume hereafter $\Sigma_{12}\ne 0$, as \eqref{eq:ineg_for_global} trivially holds otherwise.

\paragraph{Step 1: $Z_1^2\indep F|V$.} 
\label{par:_z_1_indep_f_v}

Recall from \eqref{eq:no_null_v} that $\Sigma_{12}Z_2\ne 0$ a.s.. Then, let $\eps:=\Sigma_{12}Z_2/|\Sigma_{12}Z_2|$, so that $\eps\in\{-1,1\}$, and $N:=Z_1-\Sigma_{12}Z_2=Z_1-V^{1/2}\eps$. Then, $N\sim\mathcal{N}(0,1-\rho^2)$ is independent of $Z_2$, and thus also of $(V,\eps,F)$. As a result,
\begin{align*}
& P(Z_1^2\le x|V=v,F=y) \\
= & P(|v^{1/2}\eps + N|\le x^{1/2}|V=v,F=y)   \\
= & P(\eps=1|V=v,F=y) P(|v^{1/2} + N|\le x^{1/2}|V=v,F=y,\eps=1) \\
+ & P(\eps=-1|V=v,F=y) P(|-v^{1/2} + N|\le x^{1/2}|V=v,F=y,\eps=-1)  \\
= & P(\eps=1|V=v,F=y) P(|v^{1/2} + N|\le x^{1/2}) \\
+ & P(\eps=-1|V=v,F=y) P(|-v^{1/2} + N|\le x^{1/2})  \\
= & P(\eps=1|V=v,F=y) P(|v^{1/2} + N|\le x^{1/2}) \\
+ & P(\eps=-1|V=v,F=y) P(|v^{1/2} + N|\le x^{1/2})  \\
= & P(|v^{1/2} + N|\le x^{1/2}),
\end{align*}
where the third equality follows because $N$ is independent of $(V,\eps,F)$ and the fourth because $N$ has a symmetric distribution. With the same reasoning, we obtain that $P(Z_1^2\le x|V=v)=P(|v^{1/2} + N|\le x^{1/2})$. The result follows.


\paragraph{Step 2: $g_x$ is decreasing.} 
\label{par:step_2_g_x_is_decreasing}

By Anderson's theorem, $v\mapsto P(|v^{1/2} + N|\le x^{1/2})$ is non-increasing. The result follows since, by what precedes, $g_x(v)=P(|v^{1/2} + N|\le x^{1/2})$.


\paragraph{Step 3: $h_y$ is decreasing.} 
\label{par:step_3_h_y_is_decreasing}

Let $O$ denote an orthogonal matrix with first line equal to $\Sigma_{12}/\rho$ (recall that $\rho>0$ since $\Sigma_{12}\ne 0$) and let $O_{-1}$ denote its submatrix excluding the first line. Let $\widetilde{Z}_2:=O_{-1}Z_2$, so that $\widetilde{Z}_2$ is independent of $V$. Then, since
$$F=Z_2'Z_2=(OZ_2)'(OZ_2),$$
we have $F= V/\rho^2 + \|\widetilde{Z}_2\|^2$. As a result, for all $v$,
$$h_y(v) = P(V/\rho^2 + \|\widetilde{Z}_2\|^2\le y|V= v) = P(\|\widetilde{Z}_2\|^2\le y - v/\rho^2),$$
which is decreasing in $v$. The result follows.


\paragraph{Step 4: conclusion.} 
\label{par:step_4_conclusion}

We have
\begin{align*}
	P(Z_1^2\le x, F\le y)  & = E\left[E\left[\ind{Z_1^2\le x}\ind{F\le y}|V\right]\right] \\
	 & = E\left[E\left[\ind{Z_1^2\le x}|V\right] E\left[\ind{F\le y}|V\right]\right] \\
	 & =  E\left[g_x(V)h_y(V)\right] \\
	 & \ge E[g_x(V)]E[h_y(V)] \\
	 & = P(Z_1^2\le x)P(F\le y).
\end{align*}
where the first equality follows by Step 1 and the inequality holds by Steps 2 and 3 and Chebyshev's integral inequality. Hence, Equation \eqref{eq:ineg_for_global} and thus the theorem holds.



\subsection{Details on the arguably exogenous example} 
\label{sub:details_on_example_ae}

\subsubsection{Verification of the conditions of Theorem \ref{thm:altern_glob}} 
\label{ssub:verification_of_theorem_alt_glob}

It suffices to show that $\mu_1(\delta)=\Sigma_{12}\mu_2(\delta)$ for all $\delta$. Recall that $L[Y_n(0)|X_{b,n},D,X_{b,n}D]=\alpha_0 + \alpha_1'X_{b,n}$. Combined with \eqref{eq:DGP_AE_alt}-\eqref{eq:DGP_AE_alt2}, this yields
\begin{align}
L[\widetilde{Y}(0)|\widetilde{X}_b,D,\widetilde{X}_bD]=& \alpha_0 + \alpha_1'X_{b,n} - D\eta_{0n} \notag \\
= & \alpha_0 + \alpha_1'(\widetilde{X}_b + D\theta_{0n}) - D\eta_{0n} \notag \\
= & \alpha_0 + \alpha_1'\widetilde{X}_b + D\left(\alpha_1'\theta_{0n} - \eta_{0n}\right).\label{eq:BLP_Ytilde}
\end{align}
Because the distribution of $(D,\widetilde{X}_b,\widetilde{Y}(0))$ does not depend on $n$, $\alpha_1'\theta_{0n} - \eta_{0n}=c$, a constant independent of $n$. By letting $n\to\infty$, we obtain $c=0$. Theorefore, $\alpha_1'\delta_{\theta}=\delta_\eta$. Because $\mu_1(\delta)=\Sigma_\beta^{-1/2}\delta_\eta$ and $\mu_2(\delta)=\Sigma_\theta^{-1/2}\delta_\theta$, we get $\mu_1(\delta)= \Sigma_\beta^{-1/2} \alpha_1'\Sigma_\theta^{1/2}\mu_2(\delta)$. Hence, it suffices to prove
\begin{equation}
\Sigma_{12}=\Sigma_\beta^{-1/2} \alpha_1'\Sigma_\theta^{1/2}.	
	\label{eq:eq_for_appli_Thm4}
\end{equation}

\medskip
For any random variable $A_i$, let $\overline{A}^d$ denote its average in the subsample $\{D_i=d\}$, and let $\widetilde{\text{TE}}_i=\widetilde{Y}_i(1)-\widetilde{Y}_i(0)$; remark that by \eqref{eq:DGP_AE_alt}-\eqref{eq:DGP_AE_alt2}, $Y_{n,i}(1)-Y_{n,i}(0)=\widetilde{\text{TE}}_i - D_i \eta_{n0}$. Then,
\begin{align*}
	\widehat{\beta} = & \overline{Y_n(1)-Y_n(0)}^1 + \overline{Y_n(0)}^1 - \overline{Y_n(0)}^0 =\overline{\widetilde{\text{TE}}}^1 - \eta_{n0} + \alpha_1' \widehat{\theta} - \overline{\eps_n}^1 - \overline{\eps_n}^0,
\end{align*}
where $\eps_{n,i}:=Y_{n,i}(0) - \alpha_0 - \alpha_1'X_{b,n,i}$. It follows from \eqref{eq:DGP_AE_alt}-\eqref{eq:DGP_AE_alt2} and $\alpha_1'\theta_{0n} = \eta_{0n}$ that $\eps_{n,i}=\widetilde{Y}_i(0) - \alpha_0 - \alpha_1'\widetilde{X}_{b,i}$ and thus does not depend on $n$; accordingly, we now denote it by $\eps_{i}$.

\medskip
Let $\psi_\beta$ and $\psi_\theta$ denote the influence functions of $\widehat{\beta}$ and $\widehat{\theta}$ respectively. We have
\begin{align}
	\psi_\beta & = \frac{D}{P(D=1)}\left[\widetilde{\text{TE}}-E[\widetilde{\text{TE}}|D=1]) + \eps-E[\eps|D=1]\right] + \alpha_1'\psi_\theta - \frac{(1-D)(\eps-E[\eps|D=0])}{P(D=0)}, \notag \\
	\psi_\theta & = \frac{D}{P(D=1)}(\widetilde{X}_b-E[\widetilde{X}_b|D=1]) - \frac{1-D}{P(D=0)}(\widetilde{X}_b-E[\widetilde{X}_b|D=0]).\label{eq:psi_theta}
\end{align}
Then, the asymptotic covariance $C_a$ between $\widehat{\beta}$ and $\widehat{\theta}$ satisfies $C_a=\Cov(\psi_\beta,\psi_\theta)$. Conditions \eqref{eq:linear_model} and \eqref{eq:nocov_effects_X} imply that $\Cov(\eps,D)=0$, $\Cov(\eps,D\widetilde{X}_b)=\Cov(\eps,\widetilde{X}_b)=0$ and $\Cov(\widetilde{\text{TE}},\widetilde{X}_b|D=1)=0$. As a result,
$$C_a=\alpha_1'V(\psi_\theta) = \alpha_1' \Sigma_\theta.$$
Because $\Sigma_{12}= \Sigma_\beta^{-1/2} C_a \Sigma_\theta^{-1/2}$, we obtain \eqref{eq:eq_for_appli_Thm4}.


\subsubsection{Verification of the conditions of Numerical Result \ref{num:altern_almost_glob}.} 
\label{ssub:verification_of_numerical_result}

As in \eqref{eq:BLP_Ytilde}, we have
$$L[\widetilde{Y}(0)|\widetilde{X}_b,,D,\widetilde{X}_bD,\widetilde{W}]= \alpha_0 + \alpha_1'\widetilde{X}_b + \widetilde{W} + D\left(\alpha_1'\theta_{0n} - \eta_{0n} + \tau_n\right).$$
This implies $\tau_n = \eta_{0n} - \alpha_1'\theta_{0n}$. As a result, $n^{1/2}\tau_n$ admits a limit, $\delta_\tau$ say, that satisfies $\delta_\tau=\delta_\eta-\alpha_1'\delta_\theta$.

\medskip
Let $\eps_{n,i}:=Y_{n,i}(0) - \alpha_0 - \alpha_1'X_{b,n,i} - Z_{n,i}$. Then, also, $\eps_{n,i}:=\widetilde{Y}_i(0) - \alpha_0 - \alpha_1'\widetilde{X}_{b,i} - \widetilde{W}_{i}$, which does not depend on $n$. Using the same notation as above, we have
$$\widehat{\beta} = \overline{\widetilde{\text{TE}}}^1 - \eta_{n0} + \alpha_1' \widehat{\theta} + \overline{\widetilde{W}}^1 - \overline{\widetilde{W}}^0 + \tau_n + \overline{\eps_n}^1 - \overline{\eps_n}^0.$$
The influence function of $\widehat{\beta}$ is thus as above, up to the terms corresponding to $\overline{\widetilde{W}}^1 - \overline{\widetilde{W}}^0$.

\medskip
Now, let us consider the influence function of  $\widehat{\alpha}_1'\widehat{\theta}$, where we recall that $\widehat{\alpha}_1$ is the coefficient of $X_{b,n}$ from a regression of $Y_{i,n}$ on $(1,X_{b,i,n}')$ in the untreated sample. Given \eqref{eq:linear_model_OVB}, $X_{b,n}=\widetilde{X}_b+D\theta_{0n}$, $W_n=\widetilde{W}+D\tau_n$ and $\Cov(\widetilde{X}_b,\widetilde{W}|D)=0$, $\widehat{\alpha}_1$ is root-$n$ consistent for $\alpha_1$. Then,
$$n^{1/2}\widehat{\alpha}_1'\widehat{\theta}=n^{1/2}\alpha_1'\widehat{\theta} + n^{1/2}(\widehat{\alpha}_1-\alpha_1)'\widehat{\theta}  = n^{1/2}\alpha_1'\widehat{\theta} + o_P(1),$$
where the second equality follows because $\widehat{\theta}=o_P(1)$. Hence, the influence function of $\widehat{\alpha}_1'\widehat{\theta}$ is $\alpha_1'\psi_\theta$, with $\psi_\theta$ given by \eqref{eq:psi_theta}.

\medskip
As a result, the asymptotic covariance $C_a$ between $\widehat{\beta}$ and $\widehat{\alpha}_1'\widehat{\theta}$ satisfies $C_a=V(\alpha_1'\psi_\theta)$. Because $\Sigma_{12}= \Sigma_\beta^{-1/2} C_a V(\alpha_1'\psi_\theta)^{-1/2}$, we obtain $\Sigma_{12}= \Sigma_\beta^{-1/2} V(\alpha_1'\psi_\theta)^{1/2}$. Next, as above, $\mu_2(\delta)= V(\alpha_1'\psi_\theta)^{-1/2}\alpha_1'\delta_\theta$ and $\mu_1(\delta)= \Sigma_\beta^{-1/2}\delta_\eta = (1+R(\delta))\Sigma_\beta^{-1/2}\alpha_1'\delta_\theta$, with $R(\delta)=\delta_\tau/\alpha_1'\delta_\theta$. Hence, $\mu_1(\delta)=(1+R(\delta))\Sigma_{12}\mu_2(\delta)$, as required.

\subsection{Details on Example DID with differential monotone trends} 
\label{sub:details_on_example_did}

As mentioned below Proposition \ref{prop:conserv}, the asymptotic covariance $C$ between $\widehat{\beta}$ and $\widehat{\theta}$ satisfies
\begin{equation}
C=\left(\frac{1}{P(G=1)}+\frac{1}{P(G=0)}\right)V(\eps_{t_0-1})(1-\rho^\ell)\left[1-\rho^{t_0-2},....,1-\rho\right].	
	\label{eq:cov_DID}
\end{equation}
Also, $\eta_{0n}=\lambda_{n,t_0-1+\ell}$ and $\theta_{0n}=[\lambda_{n,1},...,\lambda_{n,t_0-1}]$, implying that $\delta_\eta = K e_{t_0-2}' \delta_\theta$, with $K:=\lambda_{n,t_0-1+\ell}/\lambda_{n,t_0-2}<0$ (by monotonicity) and $e_{t_0-2}$ the $t_0-2$th canonical vector in $\R^{t_0-2}$. Note also that $\Sigma_{12}=\Sigma_\beta^{-1/2} C \Sigma_\theta^{-1/2}$. Now, suppose that $\mu_1(\delta)=\Sigma_{12}\mu_2(\delta)$ for all $\delta$. Since $\mu_1(\delta)=\Sigma_\beta^{-1/2}\delta_\eta$ and $\mu_2(\delta)=\Sigma_\theta^{-1/2}\delta_\theta$, we get
$$K e_{t_0-2}' \delta_\theta = \delta_\eta = C \Sigma_\theta^{-1} \delta_\theta.$$
Since this must hold for all $\delta_\theta$, we obtain $\Sigma_{\theta} e_{t_0-2} = C'/K$. Now, similar computations as those yielding \eqref{eq:cov_DID} show that the $(t,t')$-th term of $\Sigma_\theta$ satisfies
\begin{align*}
\Sigma_{\theta,t,t'} = & \left(\frac{1}{P(G=1)}+\frac{1}{P(G=0)}\right)V(\eps_{t_0-1})\left[\rho^{|t-t'|}+1 - \rho^{t_0-1-t} - \rho^{t_0-1-t'}\right] \\
= & \left(\frac{1}{P(G=1)}+\frac{1}{P(G=0)}\right)V(\eps_{t_0-1})\left(1+\rho^{|t-t'|}\right)\left(1-\rho^{t_0-1-\max(t,t')}\right)>0.	
\end{align*}
As $\Sigma_{\theta} e_{t_0-2}$ is the last column of $\Sigma_{\theta}$, all its terms are positive. We reach a contradiction, since all terms of $C'/K$ are negative, in view of \eqref{eq:cov_DID} and $K<0$.



\section{Further material on the calibration exercise}

\renewcommand{\arraystretch}{0.95}
\input{tables/tab_details_extremes}

\end{document}

%% file: tables/tab_details_center.tex
\begin{table}[H]
\centering
\begin{threeparttable}
\caption{UC \& CC, with DGPs calibrated to 12 DID articles, assuming differential linear trends}
\label{tab:details_center}
\begin{tabular}{lccccc}
\toprule
\textbf{Paper} & $\mu_2(\delta)$ & $\Sigma_{12}$ & UC & CC & CC/UC \\
& (1) & (2) & (3) & (4) & (5) \\
\midrule
\cite{bailey2015} & -1.67 & 0.24 & 68.2 & 62.8 & 0.921 \\
\cite{he2017} & -0.88 & 0.25 & 85.0 & 83.8 & 0.986 \\
\cite{markevich2018} & -0.85 & 0.42 & 80.2 & 78.1 & 0.974 \\
\cite{tewari2014} & -0.82 & 0.25 & 84.9 & 83.8 & 0.988 \\
\cite{deryugina2017} & -0.66 & 0.39 & 90.9 & 90.4 & 0.995 \\
\cite{ujhelyi2014} & -0.46 & 0.20 & 94.6 & 94.6 & 1.000 \\
\cite{kuziemko2014} & 0.39 & 0.39 & 94.3 & 94.6 & 1.003 \\
\cite{fitzpatrick2014} & 0.42 & 0.15 & 93.3 & 93.2 & 0.999 \\
\cite{lafortune2017} & 0.51 & -0.15 & 86.7 & 87.2 & 1.006 \\
\cite{deschenes2017} & 0.54 & 0.06 & 93.6 & 93.6 & 0.999 \\
\cite{gallagher2014} & 1.54 & 0.66 & 65.6 & 52.6 & 0.801 \\
\cite{bosch2014} & 2.31 & -0.03 & 34.3 & 35.3 & 1.031 \\
\midrule
\textbf{Average of the absolute value} & 0.92 & 0.27 & 81.0 & 79.2 & 0.975 \\
\bottomrule
\end{tabular}
\begin{tablenotes}
\footnotesize
\item \textit{Notes:} For the 12 DID papers in the meta-analysis of \cite{roth2022pretest}, Columns (1) and (2) of the table give $\mu_2(\delta)$, the t-statistic of the first pre-trend estimate, and $\Sigma_{12}$, the correlation between the pre-trend and event-study estimates. Then, Columns (3) and (4) give UC and CC, the unconditional and conditional coverage rates of the CI of the first event-study estimate, under the assumption that $\mu_1(\delta)=-(\Sigma_{\theta}/\Sigma_{\beta})^{1/2}\mu_2(\delta)$. Column (5) gives CC/UC.
\end{tablenotes}
\end{threeparttable}
\end{table}

%% file: tables/tab_details_extremes.tex
\begin{table}[H]
\centering
\begin{threeparttable}
\caption{UC \& CC, with DGPs calibrated to 12 DID articles, assuming differential linear trends}
\label{tab:details_extremes}
\small
\begin{tabular}{lccccc}
\toprule
\textbf{Paper} & $\mu_2(\delta)$ & $\Sigma_{12}$ & UC & CC & CC/UC \\
& (1) & (2) & (3) & (4) & (5) \\
\midrule
\multicolumn{6}{l}{\textit{Panel A: $\mu_2(\delta)$ calibrated to t-stat of pre-trend $-$ 1.96}} \\
\cite{bailey2015} & -3.63 & 0.24 & 10.2 & 3.4 & 0.332 \\
\cite{he2017} & -2.84 & 0.25 & 16.2 & 8.4 & 0.515 \\
\cite{markevich2018} & -2.81 & 0.42 & 4.5 & 0.7 & 0.146 \\
\cite{tewari2014} & -2.78 & 0.25 & 12.5 & 6.2 & 0.501 \\
\cite{deryugina2017} & -2.62 & 0.39 & 34.4 & 17.2 & 0.500 \\
\cite{ujhelyi2014} & -2.42 & 0.20 & 83.1 & 77.2 & 0.929 \\
\cite{kuziemko2014} & -1.57 & 0.39 & 83.6 & 78.7 & 0.942 \\
\cite{fitzpatrick2014} & -1.54 & 0.15 & 69.9 & 67.2 & 0.961 \\
\cite{lafortune2017} & -1.45 & -0.15 & 34.2 & 37.0 & 1.081 \\
\cite{deschenes2017} & -1.42 & 0.06 & 85.3 & 84.7 & 0.993 \\
\cite{gallagher2014} & -0.42 & 0.66 & 92.9 & 93.6 & 1.007 \\
\cite{bosch2014} & 0.35 & -0.03 & 93.5 & 93.5 & 1.000 \\
\midrule
\multicolumn{6}{l}{\textit{Panel B: $\mu_2(\delta)$ calibrated to t-stat of pre-trend $+$ 1.96}} \\
\cite{bailey2015} & 0.29 & 0.24 & 94.3 & 94.3 & 1.001 \\
\cite{he2017} & 1.08 & 0.25 & 80.0 & 78.0 & 0.975 \\
\cite{markevich2018} & 1.11 & 0.42 & 69.7 & 65.1 & 0.935 \\
\cite{tewari2014} & 1.14 & 0.25 & 75.3 & 72.8 & 0.966 \\
\cite{deryugina2017} & 1.30 & 0.39 & 78.3 & 74.0 & 0.945 \\
\cite{ujhelyi2014} & 1.50 & 0.20 & 90.6 & 89.1 & 0.984 \\
\cite{kuziemko2014} & 2.35 & 0.39 & 68.9 & 53.3 & 0.773 \\
\cite{fitzpatrick2014} & 2.38 & 0.15 & 40.0 & 33.9 & 0.849 \\
\cite{lafortune2017} & 2.47 & -0.15 & 1.9 & 2.8 & 1.461 \\
\cite{deschenes2017} & 2.50 & 0.06 & 64.3 & 61.7 & 0.961 \\
\cite{gallagher2014} & 3.50 & 0.66 & 5.7 & 0.0 & 0.002 \\
\cite{bosch2014} & 4.27 & -0.03 & 0.8 & 1.0 & 1.216 \\
\midrule
\textbf{Average of the absolute value} & 1.99 & 0.27 & 53.8 & 49.7 & 0.832 \\
\bottomrule
\end{tabular}
\begin{tablenotes}
\item \textit{Notes:} see notes of Table \ref{tab:details_center}.
\end{tablenotes}
\end{threeparttable}
\end{table}